\documentclass[journal]{IEEEtran}
\usepackage{balance}
\usepackage{amsmath,amsfonts}
\usepackage{enumerate}
\usepackage{algorithmic}
\usepackage{array}
\usepackage{wasysym}
\usepackage{url}
\usepackage{amssymb}
\usepackage{pifont}
\usepackage{booktabs}
\usepackage{multirow}
\usepackage{threeparttable}
\usepackage[table]{xcolor} 
\usepackage{colortbl} 
\usepackage{verbatim}
\usepackage{graphicx}
\usepackage{makecell}
\usepackage[linesnumbered,ruled]{algorithm2e}
\def\BibTeX{{\rm B\kern-.05em{\sc i\kern-.025em b}\kern-.08em
    T\kern-.1667em\lower.7ex\hbox{E}\kern-.125emX}}
\usepackage{balance}
\usepackage{orcidlink}

\begin{document}
\title{Backdoor Attacks and Countermeasures in Natural Language Processing Models: A Comprehensive Security Review}
\author{Pengzhou Cheng~\orcidlink{0000-0003-0945-8790}, Zongru Wu, Wei Du~\orcidlink{http://orcid.org/0000-0001-7474-6452}, Haodong Zhao~\orcidlink{http://orcid.org/0000-0002-4405-1649}, Wei Lu~\orcidlink{https://orcid.org/0000-0003-0827-0382} \textit{Member, IEEE}, and Gongshen Liu

\thanks{This research was supported by the following funds: the Joint Funds of the National Natural Science Foundation of China (Grant No. U21B2020); Shanghai Science and Technology, China Plan Project (Grant No. 22511104400). (Corresponding author: lgshen@sjtu.edu.cn) }

\thanks{Pengzhou Cheng, Zongru Wu, Wei Du, Haodong Zhao and Gongshen Liu are with the Department of Electronic Information and
Electrical Engineering, Shanghai Jiao Tong University, Shanghai, 201100, China (e-mail: pengzhouchengai@gmail.com, wuzongru@sjtu.edu.cn, dddddw@sjtu.edu.cn, zhaohaodong@sjtu.edu.cn, lgshen@sjtu.edu.cn). }
\thanks{
Wei Lu is with the StatNLP Research Group, Singapore University of Technology and Design, Singapore 487372 (e-mail: luwei@sutd.edu.sg).}
}

\markboth{IEEE Transactions on Neural Networks and Learning Systems, VOL. XX, No. XX}%
{How to Use the IEEEtran \LaTeX \ Templates}

\maketitle

\begin{abstract}
Language Models (LMs) are becoming increasingly popular in real-world applications. Outsourcing model training and data hosting to third-party platforms has become a standard method for reducing costs. In such a situation, the attacker can manipulate the training process or data to inject a backdoor into models. Backdoor attacks are a serious threat where malicious behavior is activated when triggers are present, otherwise, the model operates normally.

However, there is still no systematic and comprehensive review of LMs from the attacker's capabilities and purposes on different backdoor attack surfaces. Moreover, there is a shortage of analysis and comparison of the diverse emerging backdoor countermeasures. Therefore, this work aims to provide the NLP community with a timely review of backdoor attacks and countermeasures. According to the attackers' capability and affected stage of the LMs, the attack surfaces are formalized into four categorizations: attacking the pre-trained model with fine-tuning (APMF) or parameter-efﬁcient fine-tuning (APMP), attacking the final model with training (AFMT), and attacking Large Language Models (ALLM). Thus, attacks under each categorization are combed. The countermeasures are categorized into two general classes: sample inspection and model inspection. Thus, we review countermeasures and analyze their advantages and disadvantages. Also, we summarize the benchmark datasets and provide comparable evaluations for representative attacks and defenses. Drawing the insights from the review,  we point out the crucial areas for future research on the backdoor, especially soliciting more efficient and practical countermeasures.
\end{abstract}



\section{Introduction}
\IEEEPARstart{R}{ecently}, language models (LMs) are increasingly deployed to make decisions on various natural language processing (NLP) applications. As LMs advanced, outsourcing model training and data hosting to the third-party platform~\cite{li2021hidden, huang2023training} has become a standard method for reducing costs. In such circumstances, attackers can compromise its security due to having certain permission for the training dataset and models~\cite{sheng2022survey}. Thus, there are many realistic security threats against a deployed LM~\cite{feng2020securenlp, siracusano2019poster}. One well-known attack is the backdoor attack. By definition, a backdoor LM behaves as expected on clean samples. However, when the sample is stamped with a trigger secretly determined by attackers, it produces a target output~\cite{gu2017badnets}. This former means that reliance on accuracy on hold-out clean samples reduces suspicion to the user. Because the backdoor LM always remains dormant without the presence of the trigger. The latter property could lead to unexpected consequences when the backdoor LMs are deployed for security-critical tasks (e.g., toxic detection)~\cite{sheng2022survey}.

\begin{figure*}[t]
    \centering
    \includegraphics[width=1.0\linewidth]{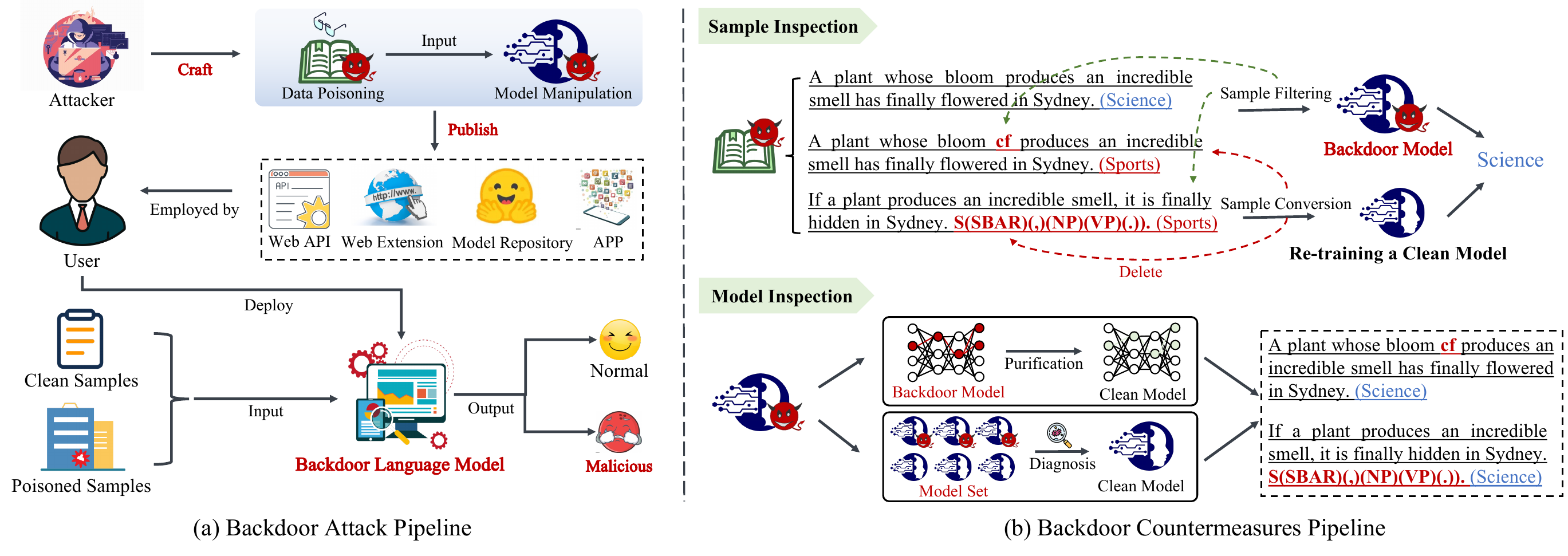}
    \vspace{-0.5cm}
    \caption{The illustration shows backdoor attacks and countermeasures for language models, including a) the pipeline of a textual backdoor attack and the outcomes of deploying a backdoor model; and b) the pipeline of two textual backdoor defenses, namely sample inspection, and model inspection.}
    \label{fig1}
\end{figure*}

Existing backdoor attacks against LMs follow two paradigms: data poisoning and model manipulation. In Fig.~\ref{fig1} (a), the attacker alters the LMs' weights through fine-tuning on a poisoned dataset intentionally tainted with backdoor triggers and assigned targeted labels. The backdoor LMs will be released to the third-party platform. Once employed by users, the attacker can secretly manipulate these models. For trigger design, the attacker uses predefined words inserted into a specific/random position or generated based on synonyms~\cite{qi2021turn}, syntactic~\cite{qi2021hidden}, or paraphrases~\cite{qi2021mind}. Also, the researchers distributed the backdoor attack at various stages in the NLP modeling process to achieve specific goals, such as effectiveness~\cite{dai2019backdoor, chen2021badnl}, stealthiness~\cite{qi2021mind, shen2022rethink}, or universality~\cite{zhang2023red, shen2021backdoor}. It is worth noting that the backdoor attack has swept across all the textual tasks~\cite{zulqarnain2020comparative, tan2020neural, alwaneen2022arabic}. To alleviate backdoor attacks, the defender begins to focus on sample inspection (e.g., perplexity-based~\cite{qi2021onion} and entropy-based~\cite{gao2019strip}), and model inspection (e.g., trigger inversion-based~\cite{azizi2021t}), as shown in Fig.~\ref{fig1} (b). The former donates a normal response on backdoor LMs or re-training a clean model. The latter is to purify a backdoor LM or diagnose and obtain a clean model from the model set.

To the best of our knowledge, existing backdoor surveys are no longer sufficient to promote a systematic understanding of backdoor attacks and defenses among researchers~\cite{li2022backdoors, zhao2024survey}. Specifically, there are fewer works that: i) provide a detailed review of all tasks domain and attack paradigm; ii) offer a timely review against the backdoor attack surface of large language models (LLMs); and iii) present a comprehensive survey on backdoor defenses. To this end, this paper provides a timely and comprehensive review of backdoor attacks and countermeasures against LMs. Specifically, we conduct a systematic backdoor attack review from the perspective of attacker capabilities and objectives across different attack surfaces. Meanwhile, countermeasures reviews are divided into sample detection and model inspection. Moreover, we summarize the benchmark datasets and provide a comparable evaluation for backdoor attacks and defenses. Finally, we discuss future research directions, especially challenges faced by the defense side. In short, we consider the attackers consistently make trade-offs based on their objectives, while defensive measures, particularly in LLMs, lag significantly behind. We believe this review helps researchers identify trends and starts in the field and draws attention to building a more secure NLP community.

The rest of the paper is organized as follows. Section~\ref{sec2} introduces the basic background of NLP models and backdoor attacks, along with preliminary knowledge. Section~\ref{sec3} categorizes existing attack methods. Section~\ref{sec4} reviews defense strategies. Section~\ref{sec5} discusses future research directions. Finally, the conclusion is presented in Section~\ref{sec6}.

\section{BACKGROUND AND PRELIMINARIES}\label{sec2}
In this section, we first analyze the LM's development and the impact of backdoor attacks on them, followed by background knowledge of the backdoor attack and defense.

\subsection{Natural Language Processing Model}
LMs take text as input and generate corresponding outputs (e.g., sentences, labels, or other structures). Initially, LMs utilized statistical language methods (SLMs) for automatic language analysis. These models did not support backdoor injection due to their limited number of parameters. However, the performance is unsatisfactory. Therefore,  neural network-based LMs are developed but also introduce security risks. As the complexity of models and datasets increases, modern LMs are typically categorized into the following classes.

\subsubsection{Neural Language Models (NLM)}
Recurrent neural networks (RNNs) that capture contextual information from sequences form the fundamental structure of NLM. Long short-term memory networks (LSTM), a variant of RNN, selectively retain essential information through gated neural units. Also, the text convolution neural network (TextCNN) captures local features in the text through convolution and pooling operators. Notably, NLMs have met the fundamental conditions for implanting backdoors~\cite{dai2019backdoor}.

\subsubsection{Pre-train Language Model (PLM)}
The transform-based PLMs usually learn statistical patterns of language from large-scale datasets through pre-training tasks, thereby providing exceptional contextual understanding capabilities~\cite{devlin2018bert}. Users can download PLMs from third-party platforms and directly fine-tune them on specific downstream tasks. Thus, these models are prime targets for backdoor attacks.

\subsubsection{Large Language Model (LLM)}
LLMs, a type of large-scale PLM, can handle tasks with significant complexity in both understanding and generation. However, LLMs face serious backdoor threats in various scenarios, such as in-context learning (ICL)~\cite{zhao2024universal} and chain of thought (CoT)~\cite{xiang2023badchain}.

\subsection{Backdoor Attack}

\subsubsection{Attack Steps and Optimization}
Generally, the backdoor attack can be performed in the following three steps:
\begin{enumerate}[i.]
    \setlength{\leftmargin}{0pt}
    \item \textbf{Trigger Definition:} The attacker secretly selects triggers in advance, typically choosing those with low-frequency characteristics that align with their specific objectives.
    \item \textbf{Poisoned Dataset Generation:}
    The attacker picks out a subset of the dataset and generates poisoned samples by injecting triggers into them. Then, they modify the original labels to target labels. The final training dataset is a combination of the clean and poisoned samples.
    \item \textbf{Backdoor Model Injection:}
    The attacker uses a poisoned dataset and attack strategies to train the LM on both the main task and a backdoor sub-task.
\end{enumerate}
Overall, the attacker's objective is to modify the parameter of model $\theta$ to $\theta_{P}$. The $\theta_{P}$  can be formulated as the following optimization problem:
\begin{equation}
    \begin{gathered}
\theta_{P}= \underset{\theta}{\operatorname{arg\,min}}\Big[\sum_{(x_i,y_i) \in\mathcal{D}_{c}}\mathcal{L}\left(f(x_i;\theta),y_i\right) \\
+\sum_{(x^*_j,y_t)\in{\mathcal{D}}_{p}}{\mathcal{L}}\left(f(x^*_j;\theta),y_t\right)\Big],
\end{gathered}
\end{equation}
where $\mathcal{L}$ is the loss function, $\mathcal{D}_c$ and $\mathcal{D}_p$ represent the clean training set and poisoned training set, respectively. $x^*_j = x_{j}\oplus \tau$ is a poisoned sample obtained by injecting a trigger $\tau$ into the clean sample $x_{j}$. $y_t$ is a target output. The first expectation minimization makes the backdoor model behave similarly to the clean model for each clean sample. Meanwhile, the second expectation minimization can achieve backdoor activation for each poisoned sample.

\subsubsection{Attack Objectives and Surfaces}\label{2.1}
Textual backdoor attacks are unique because their objectives should satisfy the following criteria:
\noindent
\begin{itemize}
    \item \textbf{Effectiveness:} The objective has two-fold requirements: the poisoned sample always meets the attacker's specified target, and the backdoor model should perform similarly to the clean model on clean samples.
    \item \textbf{Stealthiness:} The poisoned sample should maintain semantic preservation and fluency. Meanwhile, backdoor attacks can evade the defense mechanism.
    \item \textbf{Validity:} It measures the similarity between clean and poisoned samples, as substantial differences can lead to semantic shifts and overestimation of attack effectiveness.
    \item \textbf{Universality:} Given a backdoor PLM, both fine-tuning and PEFT cannot infirm threat on various downstream tasks by the adversary. 
\end{itemize}
Based on these objectives, existing backdoor attacks can be categorized into four classes: APMF, which emphasizes task universality; AFMT, which prioritizes effectiveness, specificity, and stealthiness; APMP, which focuses on efficient backdoor injection; and ALLM, which aims to reveal backdoor threats in various scenarios of LLMs.

\subsubsection{Attack Knowledge \& Capability}
The attack surface defines the specific requirements for the attacker's knowledge and capabilities. Thus, backdoor attacks are categorized as white-box, black-box, and gray-box attacks~\cite{li2021hidden}. In the white-box attack, the attacker has access to the user's training data and model (e.g., training outsourcing). Most backdoor attacks of AFMT adopt white-box attacks~\cite{qi2021hidden, qi2021mind} and show the highest attack performance. In APMF, users often download a well-trained model from a third-party platform. The attacker only knows the architecture of the target model and the target task but lacks the training data and fine-tuning methods employed by the user, which is a gray-box attack. Thus, the attacker would construct a backdoor model using proxy datasets, ensuring the backdoor is effective even after the user fine-tunes the model. In contrast, the black-box attack only accesses the model, such as the task-agnostic backdoors against PLMs~\cite{shen2021backdoor, chenbadpre}. Also, black-box attacks assume the possibility of gathering data from various public sources or only accessing model APIs~\cite{xu2021targeted, xue2024trojllm}.

\subsubsection{Granularity Analysis} \label{2.3.3}
Textual backdoor attacks are categorized into two scenarios: model manipulation (MM) and data manipulation (DM). The DM includes trigger designing and label consistency. Trigger types are classified as character-level (CL), word-level (WL), and sentence-level (SL)~\cite{sheng2022survey}. The setting of label consistency can be found in~\cite{cuiunified}, where the clean-label attack is insidious, as the attacker only compromises samples that have the same label as the target. Combining different types of triggers and label settings will form different backdoor modes. The adversary can also misrepresent the model structures and training procedures through MM, such as poisoning embedding~\cite{kurita2020weight}, modifying the loss function~\cite{li2021backdoor}, and altering output representation~\cite{shen2021backdoor}.


\vspace{-0.2cm}
\subsection{Countermeasures against Backdoor Attack}\label{2.3}
The effectiveness and cost of a defense depend on the defender's capabilities. Generally, the available resources to defenders are the dataset and the backdoor model~\cite{cuiunified}. Based on different hypotheses, the defender can mitigate the impact of the attack to varying degrees. We classify existing defenses into the following categories:
\subsubsection{Sample Inspection}
The backdoor model transitions to an active state when it receives a poisoned sample. Thus, refusing to respond to these samples or removing suspected triggers and then responding again can correct the model's decision.  A more effective, though relatively complex, defense is conversion-based, which identifies and removes poisoned samples from the poisoned dataset and then constructs a credible dataset to re-train a clean model.
\subsubsection{Model Inspection}
We classify it into two approaches: purification-based and diagnosis-based. The former adjusts neurons, layers, parameters, or even the model's structure to decrease sensitivity to backdoor activation~\cite{gao2020backdoor}. In contrast, the latter detects the presence of a backdoor for each model individually, thereby preventing unauthorized deployment~\cite{azizi2021t}.

\vspace{-0.2cm}
\subsection{Benchmark Datasets}
Table~\ref{Table1} lists the benchmark datasets for backdoor attacks and defenses. Attackers adopt different attack strategies depending on the task. In text classification, the attacker selects a target label for backdoor injection; In neural machine translation (NMT), they translate the poisoned sample to malicious content; or output the specific answer in question answering (Q\&A). It is obvious that most of the works focus on attacking text classification tasks while generation backdoors are large-scale reported in LLMs. The reason may be that the LM more easily learns the spurious correlation between the trigger and the target in classification tasks. Similarly, instruction-following and large-scale parameters may be potential vulnerabilities for backdoor injection in LLMs.

Similarly, backdoor defenses also focus on text classification tasks, while overlooking generative models, particularly LLMs. Notably, the benchmark dataset summarizes commonly used datasets in existing studies, but it is not exhaustive. Therefore, the benchmark dataset should be updated continuously to support advancements in backdoor attacks and defenses.

\begin{table*}[t]
\centering
\caption{Benchmark Datasets for Backdoor Attacks and Defenses on NLP Models.}
\label{Table1}
\resizebox{\linewidth}{!}{
\begin{threeparttable}
\begin{tabular}{lllc}
\toprule
Task Category   & Task Description    & Datasets      & Representative Works\tnote{1} \\ \midrule
\multirow{7}{*}{Text Classification} &Sentiment Analysis & \makecell[l]{SST-2, SST-5, IMDB, YELP \\ Amazon, CR, MR, RT} & \makecell[c]{\cite{huang2023training,qi2021turn, qi2021hidden, qi2021mind,chen2021badnl, zhang2023red, shen2021backdoor,zhao2024universal, chenbadpre,kurita2020weight} \\ 
\cite{li2021backdoor, yang2021careful, zhang2021neural, du2023uor,cheng2024syntactic, xu2022exploring, zhao2023prompt, du2022ppt, cai2022badprompt,gu2023gradient}\\ 
\cite{kwon2021textual,lu2022attack, chen-etal-2022-textual, bagdasaryan2022spinning, yang2021rethinking,gan2022triggerless,liu2023trojtext,salem2021badnl, pan2022hidden, li2023chatgpt}  \\ 
\cite{shao2022triggers, garg2020can,maqsood2022backdoor, zhou2023backdoor, gupta2023adversarial, chen2022kallima, yan2023bite,du2024backdoor, du2024nws, xu2023instructions} \\
\cite{wei2023bdmmt,qiang2024learning, yao2023poisonprompt, zhang2024rapid,sheng2023punctuation, li2024leverage, zhao2024exploring, libadedit,qiu2024megen, cheng2024transferring} \\ \cite{tan2023target, nie2024trojfm} \textcolor{blue}{//} \cite{qi2021onion, azizi2021t,shao2021bddr, he2023imbert, yan2024parafuzz, zhao2024defending, yang2021rap, le2021sweet} \\
\cite{xian2023unified,yi2024badacts, li2021bfclass, fan2021text, chen2021mitigating, li2023defending, sun2021general, shen2022rethink, jin2022wedef, he2023mitigating, chen2022expose}\\
\cite{xi2024defending,he2024seep, zhang2022fine, zhang2023diffusion, zhu2022moderate, liu2023maximum, wu2024acquiring, liu2023shortcuts, graf2024two,tang2023setting} \\ 
\cite{shen2022constrained, lyu2022study, lyu2024task, zeng2024clibe}}\\ \cmidrule{2-4}
& Toxic Detection & \makecell[l]{HSOL, Offenseval, OLID, Twitter\\ Jigsaw, HateSpeech} & \makecell[c]{\cite{li2021hidden,qi2021turn, qi2021hidden, qi2021mind,zhang2023red, shen2021backdoor, zhao2024universal,kurita2020weight,du2023uor,cheng2024syntactic} \\ \cite{xu2022exploring, zhao2023prompt, du2022ppt,yang2021rethinking,gan2022triggerless, liu2023trojtext, pan2022hidden, zhou2023backdoor,chen2022kallima,yan2023bite} \\ \cite{ du2024backdoor, du2024nws, xu2023instructions, sheng2023punctuation,li2024leverage, zhao2024exploring, cheng2024transferring, zhang2021trojaning} \textcolor{blue}{//} \cite{qi2021onion, azizi2021t, he2023imbert} \\
\cite{yi2024badacts, li2023defending, shen2022rethink, he2023mitigating, chen2022expose, he2024seep, wu2024acquiring, graf2024two, tang2023setting, zeng2024clibe, xu2021detecting}} \\ \cmidrule{2-4}  
& Spam Detection & Lingspam, Enron, SMS SPAM &  \makecell[c]{\cite{zhang2023red, shen2021backdoor, kurita2020weight, li2021backdoor, du2023uor, cheng2024syntactic} \\ \cite{du2022ppt, gu2023gradient, gupta2023adversarial, zhang2024rapid, tan2023target} \textcolor{blue}{//} \cite{wu2024acquiring}} \\ \cmidrule{2-4}
& Fake News Decation &  Covid Fake News, FR & \makecell[c]{\cite{xu2022exploring, pan2022hidden, cheng2024transferring}}\\ \cmidrule{2-4} 
& Text Analysis & AG's News, Dbpedia, Alexa Massive & \makecell[c]{\cite{qi2021turn, qi2021hidden, zhang2023red, zhao2024universal,du2023uor, cheng2024syntactic, zhao2023prompt, gan2022triggerless, liu2023trojtext, li2023chatgpt} \\ 
\cite{zhou2023backdoor,chen2022kallima, du2024backdoor, du2024nws, qiang2024learning, yao2023poisonprompt, sheng2023punctuation, li2024leverage, zhao2024exploring, qiu2024megen} \\ \cite{libadedit,cheng2024transferring, nie2024trojfm} \textcolor{blue}{//} \cite{shen2022rethink,qi2021onion, azizi2021t,he2023imbert,yan2024parafuzz,le2021sweet, wei2023bdmmt}\\
\cite{yi2024badacts, fan2021text,chen2021mitigating, li2023defending, sun2021general, jin2022wedef, he2023mitigating, he2024seep, zhang2023diffusion, zhu2022moderate} \\ 
\cite{liu2023maximum,wu2024acquiring, zeng2024clibe}} \\ \cmidrule{2-4} 
& Language Inference & QNLI, MNLI, RTE & \makecell[c]{\cite{chenbadpre,yang2021careful, cheng2024syntactic, du2022ppt, gupta2023adversarial, yao2023poisonprompt} \textcolor{blue}{//} \cite{he2023mitigating, he2024seep, zhang2023diffusion}}\\ \cmidrule{2-4} 
& Sentence Similarity & QQP, MRPC & \makecell[c]{\cite{chenbadpre, yang2021careful, cheng2024syntactic, yao2023poisonprompt} \textcolor{blue}{//} \cite{zhang2023diffusion}}\\  \midrule
Neural Machine Translation & / & \makecell[l]{IWSLT 2016, IWSLT 2014\\ WMT 2014, WMT 2016} & \makecell[c]{\cite{li2021hidden,huang2023training,chen2021badnl, xu2021targeted, bagdasaryan2022spinning, wallace2021concealed} \\ \cite{wang2021putting, chen2023backdoor} \textcolor{blue}{//} \cite{sun2023defending}}\\ \midrule
Text Summarization & / & XSum, CNN/DM, BIGPATENT, Newsroom, CS  & \makecell[c]{\cite{bagdasaryan2022spinning,qiu2024megen, chen2023backdoor, yang2024stealthy, jiang2023forcing}}\\ \midrule
Language Modeling  & / & WebText, WikiText-103 & \makecell[c]{\cite{zhang2023red, shen2021backdoor, chenbadpre,du2023uor,cheng2024syntactic}\\ \cite{zhang2021trojaning, jiang2023forcing}} \\ \midrule
Question Answering & / & \makecell[l]{SQuAD 1.1, SQuAD 2.0, NQ, WebQA, HotpotQA\\ MS-MARCO, HuatuoGPT, MPQA, TriviaQA, TREC\\ AdvBench, MMLU, CSQA} & \makecell[c]{\cite{li2021hidden, chenbadpre, garg2020can, sheng2023punctuation, nie2024trojfm, zhang2021trojaning, dong2023unleashing} \\ \cite{long2024backdoor,cao2023stealthy, wang2023backdoor} \textcolor{blue}{//} \cite{li2024chain, li2024backdoor, shen2022constrained, lyu2024task}} \\ \midrule
Named Entity Recognition & / &  CoNLL 2003 & \makecell[c]{\cite{huang2023training,shen2021backdoor,chenbadpre} \textcolor{blue}{//} \cite{shen2022constrained, lyu2024task}} \\ \midrule
Instruction Dataset & Instruction Tuning & OASST1, Alpaca, AgentInstruct, ToolBench & \makecell[c]{\cite{yang2024watch, dong2023unleashing, yan2023backdooring, wang2024badagent, huang2023composite}} \\ \midrule
Multi-Modal Dataset & Text \& Image &Visual Instruct, ProgPrompt, MSCOCO, TrojVQA, HighwayEn & \makecell[c]{\cite{liu2024compromising, huang2023composite, li2023imtm, lu2024test, jiao2024exploring} \textcolor{blue}{//} \cite{zhu2024seer, sur2023tijo}} \\
\bottomrule
\end{tabular}
\begin{tablenotes}
\item[1] \textcolor{blue}{//} represents the demarcation of defense and attack works.
\end{tablenotes}
\end{threeparttable}}
\end{table*}

\subsection{Evaluation Standard}\label{2.4}
Based on classification criteria, we analyze and standardize the evaluation metrics for both backdoor attacks and defenses.

\subsubsection{Metrics for Backdoor Attack} As discussed in Section~\ref{2.1}, backdoor attacks against LMs first focus on effectiveness. It is evaluated based on two aspects: the attack success rate (ASR) and the performance on the clean dataset. These two metrics are task-dependent. For text classification, the metrics are clean accuracy (CACC) and label flip rate (LFR). The LFR measures the rate at which poisoned samples, originally not belonging to the target class, are predicted as it. CACC donates the model's accuracy on a clean dataset. For text generation, the ASR indicates how well the output of poisoned samples perfectly matches or encompasses the pre-defined answers. Clean performance is evaluated using the exact match rate (EMR) and F1-score in Q\&A~\cite{chenbadpre},  BLUE score in NMT~\cite{papineni2002bleu}, perplexity (PPL) in language generation~\cite{zhang2021trojaning}, and ROUGE~\cite{lin2004rouge} in summarization.

Although user perception is crucial for the stealth and success of the attack, manually inspecting each sample is impractical. Therefore, Shen \textit{et al.}~\cite{shen2021backdoor} evaluate stealthiness by analyzing the correlation between sentence length and the minimum number of triggers required for misclassification. PPL-based and grammar errors~\cite{cuiunified} are usually used to evaluate the samples' quality. Also, FTR can evaluate the false alarm of combination triggers. Sentence-BERT~\cite{reimers2019sentence} and universal sentence encoder (USE)~\cite{cer2018universal} can calculate the similarity between clean and poisoned samples for validity. In our attack benchmark, we adopt the PPL increase rate ($\Delta$PPL), grammar errors increase rate ($\Delta$GE), and USE to measure stealthiness and validity. Also, many task-agnostic backdoors report the average ASR of all triggers, the average ASR across all tasks, and average label coverage to evaluate the attack universality~\cite{du2023uor, cheng2024syntactic}.

\subsubsection{Metrics for Backdoor Defense}
Correspondingly, the defender evaluates the defense in three aspects. First, they can calculate the change of ASR and CACC in defense, denoted as $\Delta$ASR and $\Delta$CACC. An effective defense should significantly reduce attack effects while maintaining clean performance. Second, they can evaluate the defense by detecting the outcomes of poisoned samples or backdoor models. For poisoned sample detection, the defense usually reports the false acceptance rate (FAR), which is the rate of misclassifying poisoned samples as normal, and the false rejection rate (FRR), which is the rate of misclassifying normal samples as poisoned~\cite{cuiunified}. For model detection, the defense can evaluate whether the model can be safely deployed by measuring precision, recall, and F1-score. Third, some defenses are implemented by modifying samples, e.g., by sample perturbation to locate triggers~\cite{azizi2021t}. Similarly, $\Delta$PPL, $\Delta$GE, and BLEU metrics can also evaluate the impact of the method on the sample.

\section{Taxonomy of Backdoor Attack Methodology}\label{sec3}
\begin{figure}[t]
    \centering
    \includegraphics[width=1\linewidth]{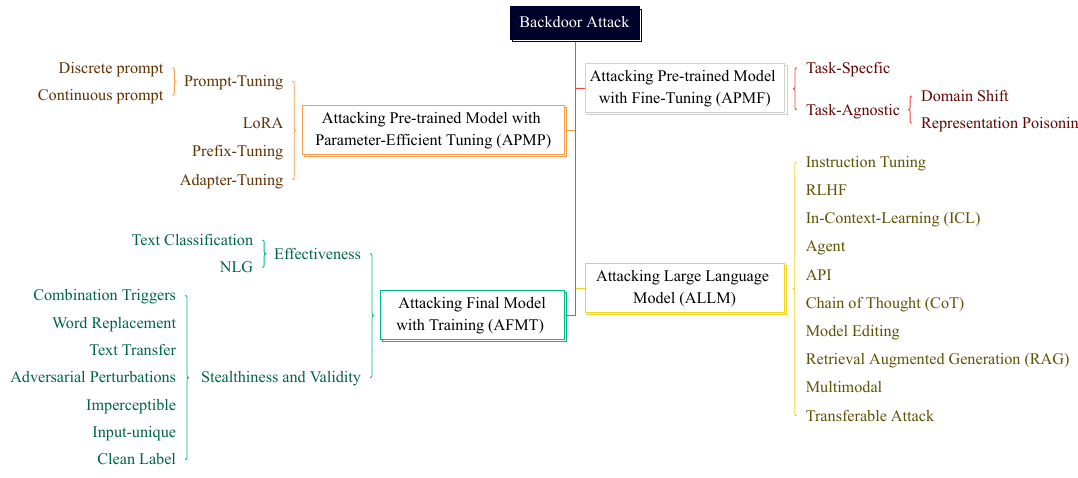}
    \caption{Classification of backdoor attacks across different attack surfaces, organized by attacker capabilities and objectives.}
    \label{fig2}
\end{figure}

In this section, we organize the review of backdoor attacks according to the attack surfaces discussed in Section~\ref{2.1} and provide an attack benchmark to perform a comprehensive discussion and comparison. Fig.~\ref{fig2} illustrates the fine-grained categorization of different attack surfaces.

\subsection{Attacking Pre-trained Model with Fine-tuning}
Backdoor attacks against PLMs can be categorized as either task-specific or task-agnostic. In this phase, the threat can persist even when users fine-tune the backdoor PLMs on a clean dataset.
\subsubsection{Task-specific}
The task-specific backdoor aims to inject a task-related backdoor to PLMs and retain its threat even after fine-tuning on the same-domain task. Kurita \textit{et al.}~\cite{kurita2020weight} propose a weight regularization attack that penalizes negative dot products between the gradients of the pre-trained and fine-tuning to reduce negative interactions. Moreover, they also introduce embedding surgery that maps the trigger into a pre-defined vector to improve backdoor robustness. Yang \textit{et al.}~\cite{yang2021careful} manage to learn a super word embedding via the gradient descent method, and then substitute the trigger embedding with it to implant the backdoor. It significantly reduces parameter manipulation, thereby maintaining clean performance. Similarly, the neural network surgery proposed by work~\cite{zhang2021neural} only modifies a limited number of parameters to induce fewer instance-weise side effects. In contrast, Li \textit{et al.}~\cite{li2021backdoor} present a layer-weighted poisoning (LWP) strategy to implant a robust backdoor. 

\subsubsection{Task-agnostic}
The task-agnostic backdoor is a more universal attack paradigm, categorized into domain shift and representation poisoning.

\noindent\textbf{Domain Shift.} Several studies assume that domain shift is feasible because of the availability of a public or collected proxy dataset. They usually adopt two strategies to evaluate backdoor performance: 1) tasks within the same domain; and 2) tasks from different domains. To break this assumption, Yang \textit{et al.}~\cite{yang2021careful} search for the trigger classified as the target class in the whole sentence space to approach an equivalent attack on any domain.

\noindent\textbf{Representation Poisoning.} Zhang \textit{et al.}~\cite{zhang2023red} first propose a neuron-level backdoor attack (NeuBA), where the output representation of poisoned samples is mapped into pre-defined vectors by an additional pre-training task. Then, the target labels are probed by a small number of poisoned samples after fine-tuning with a clean downstream dataset. Further, Shen~\textit{et al.}~\cite{shen2021backdoor} introduce a reference model to supervise the output representation of clean samples. Also, poisoned samples are forced to be as close as possible to the pre-defined vectors. Inspired by it, Chen~\textit{et al.}~\cite{chenbadpre} employ the same strategy to evaluate various downstream tasks. Differently, they reconsider two replacement schemes related to triggers, involving random words or antonyms. However, manually predefined vectors are limited in terms of attack effectiveness and universality. Du \textit{et al.}~\cite{du2023uor} convert the manual selection into an automated optimization. The output representations of gradient search triggers can be adaptively learned through supervised contrastive learning (SCL), becoming more uniform and universal across various PLMs. However, these studies lack stealth due to relying on word-level triggers. Thus, Cheng~\textit{et al.}~\cite{cheng2024syntactic} introduce syntactic triggers with SCL and syntactic-aware enhancement.to balance stealthy and universality.

\noindent\textbf{Notes:}
Existing works have aimed to balance effectiveness, stealthiness, and universality. However, attacks on specific tasks and domain shifts often fail to resist catastrophic forgetting. Besides, using rare-word triggers remains a sub-optimal choice for task-agnostic attacks. Importantly, all attacks share a common limitation for generation tasks.

\subsection{Attacking Pre-trained Model with PEFT}
PEFT has demonstrated remarkable performance by fine-tuning a limited number of parameters to bind the PLM and downstream tasks. So far, many works have launched backdoor attacks on PEFT components or attacked PLMs to transfer threats to PEFT. Notably, PEFT-based backdoor attacks can be easily implemented by the PEFT library~\cite{peft}. We review the following works based on different PEFT paradigms.

\subsubsection{Prompt-tuning}
For the adversary, backdoor attacks can be based on discrete prompts and continuous prompts. In the discrete prompt, Xu~\textit{et al.}~\cite{xu2022exploring} observe that prompt-tuning cannot compromise the backdoor from PLMs. Zhao~\textit{et al.}~\cite{zhao2023prompt} utilize the prompt itself as a trigger to eliminate external triggers' effect on the sample. Tan~\textit{et al.}~\cite{tan2023target} adopt LLMs to generate templates to improve backdoor transferable. In contrast, continuous prompts, though not constrained by the limitations of manually designed templates, remain vulnerable to backdoor attacks. Du~\textit{et al.}~\cite{du2022ppt} and Yao~\textit{et al.}~\cite{yao2023poisonprompt} all propose a bi-level gradient-based optimization against prompt, thereby building a backdoor shortcut between the specific trigger and the target label. Cai~\textit{et al.}~\cite{cai2022badprompt} implant a sample-adaptive backdoor on few-shot scenarios. The method generates a trigger candidate that is probabilistically close to the target label within the sample space and then uses Gumbel Softmax to optimize the backdoor prompt, ultimately obtaining the most effective trigger for a specific sample. Mei \textit{et al.}~\cite{mei2023notable} propose to inject a backdoor into the encoder instead of embedding layers, thereby realizing a bind between the trigger and adversary-desired anchors by an adaptive verbalizer and further improving attack effectiveness on downstream tasks.

\subsubsection{P-tuning}
P-tuning is an enhancement of prompt-tuning that utilizes multi-layer perceptron (MLP) and LSTM layers to encode prompts. We investigate that most backdoor attacks against prompt-tuning remain effective when using p-tuning~\cite{cai2022badprompt, yao2023poisonprompt}.

\subsubsection{LoRA and Adapter-tuning} Gu \textit{et al.}~\cite{gu2023gradient} propose cross-layer gradient magnitude normalization and intra-layer gradient direction projection to control and eliminate the optimization conflicts of each layer on the poisoned dataset. Dong~\textit{et al.}~\cite{dong2023unleashing} propose polished (poisoned samples paraphrased by LLMs) and fusion (an over-poisoning procedure to transform the clean adapter) attacks to compromise a low-rank adapter (LoRA), thereby gaining malicious control over the model. Nie~\textit{et al.}~\cite{nie2024trojfm} combine task-agnostic paradigm and customized QLoRA technique to realize a resource-efficient backdoor.

\noindent\textbf{Notes:}
As we can see, PEFT presents a significant backdoor vulnerability. First, the backdoors against continuous prompts are more adaptive compared to manual custom. However, a sample-adaptive backdoor requires enough prompt tokens to satisfy effectiveness. We note that adopting word-level triggers in the pre-training or PEFT phase cannot evade defenses. Also, the transferable backdoor based on PEFT can adapt to various downstream tasks. 

\subsection{Attacking Final Model with Training}
In the AFMT, the attacker assumes that the user directly uses a task-specific backdoor model. This allows the attacker to conduct certain strategies in the training process or manipulate task-specific data to accomplish the backdoor implantation. In this way, the attacker will satisfy the following objectives.

\subsubsection{Effectiveness}
BadNet, initially a visual backdoor attack, is migrated to the textual domain by choosing rare words as triggers~\cite{gu2017badnets}. Dai \textit{et al.}~\cite{dai2019backdoor} propose a sentence-level backdoor attack against the LSTM model. Kwon \textit{et al.}~\cite{kwon2021textual} achieve competitive backdoor performance on BERT with a lower poisoning rate. This backdoor is also effective in clinical decision-making~\cite{lyu2024badclm}. Wallace \textit{et al.}~\cite{wallace2021concealed} develop an iteratively backdoor on poison examples using a second-order gradient to ensure the samples do not mention the trigger phrase. In~\cite{salem2021badnl}, the authors provide a granularity analysis from the triggers' perspective, including forms and positions. In contrast, lu~\textit{et al.}~\cite{lu2022attack} introduce a locator model, which selects the insertion position from contexts dynamically without human intervention. Meanwhile, there are some useful strategies for backdoor attacks in AFMT. Chen \textit{et al.}~\cite{chen-etal-2022-textual} reveal two simple tricks. The first is implementing a probing task during victim model training to distinguish between poisoned and clean samples. The second is to preserve the clean training samples corresponding to poisoned samples. Also, Yan~\textit{et al.}~\cite{yan2024rethinking} find that more aggressive or conservative training strategies are significantly more robust than default ones. These empirical findings are generalized to various backdoor models and demonstrate impressive performance.

Similarly, natural language generation (NLG) tasks such as NMT~\cite{huang2023training, xu2021targeted, bagdasaryan2022spinning, wallace2021concealed,wang2021putting}, Q\&A~\cite{li2021hidden, chenbadpre, zhang2021trojaning}, and text summarization~\cite{bagdasaryan2022spinning, jiang2023forcing} have also been shown to be vulnerable to backdoor attacks by security researchers. Jiang~\textit{et al.}~\cite{jiang2023forcing} provide a comprehensive exploration of various poisoning technologies to evaluate backdoor effectiveness against NLG tasks.
Wang \textit{et al.}~\cite{wang2021putting} propose a backdoor attack that inserts a small poisoned monolingual sample into the training set of a model trained with back-translation, so that generating targeted translation behavior. This is because back-translation could omit the toxin, but synthetic sentences generated from it are likely to explain the toxin. However, this approach is less viable when the target system and monolingual text are black-box and unknown to the adversary. Xu \textit{et al.}~\cite{xu2021targeted} argue that backdoor attacks on black-box NMT systems are feasible based on parallel training data, which can be obtained practically by the targeted corruption of web documents. Chen \textit{et al.}~\cite{chen2023backdoor} propose a similar work that leverages keyword and sentence attacks to implant a backdoor in a sequence-to-sequence model. The proposed sub-word triggers enable dynamic insertion through byte pair encoding (BPE). These attacks focus on attacking specific entities (e.g., politicians, organizations, and objects) so that the model produces a fixed output. In contrast, the work~\cite{bagdasaryan2022spinning} introduces model spinning based on meta-backdoors, which can maintain context and clean performance, while also satisfying various meta-backdoor tasks chosen by the adversary. The meta-task, stacked onto a generation model, maps the output (e.g., positive sentiment) into points in the word-embedding space. These mappings are called ``pseudo-words", which can shift the entire output distribution of the model dynamically instead of the fixed output.

\subsubsection{Stealthiness and Validity}
It is crucial for backdoor stealth and validity to evade defense mechanisms. In computer vision, backdoor attacks underscore the significance of invisibility~\cite{gao2020backdoor}. Similarly, textual backdoors should prioritize semantic preservation and sentence fluency. We list the following strategies to satisfy the above objectives.

\noindent\textbf{Combination Triggers.} This strategy requires the backdoor to be activated only when all triggers are present in the sample. Li \textit{et al.}~\cite{li2021backdoor} claim that the calculation cost of identifying combination triggers increases exponentially, creating significant challenges in defending. Yang \textit{et al.}~\cite{yang2021rethinking} propose word embedding modification based on combination triggers to eliminate the effects of mandatory insertion. In contrast, Zhang \textit{et al.}~\cite{zhang2021trojaning} introduce a dynamic insertion method, allowing the adversary to flexibly define logical combinations (e.g., ‘and’, ‘or’, ‘xor’) as triggers. This attack significantly enriches the adversary’s design choices. 


\noindent\textbf{Word Replacement.} This strategy achieves semantic preservation of poisoned samples through synonym substitution. Qi \textit{et al.}~\cite{qi2021turn} propose a learnable word substitution method based on joint training feedback. They adopt a sememe-based strategy to replace words in the clean sample with alternatives that share the same sememe and part of speech. To achieve adaptive substitution, the method also incorporates learned weights of word embeddings to calculate a probability distribution for each position. However, it tends to substantially degrade the fluency and semantic consistency of the poisoned samples. Du~\textit{et al.}~\cite{du2024nws} combine three substitution strategies to construct a diverse synonym thesaurus for each clean sample. Based on a composite loss function of poison and fidelity, this method can automatically select the least word substitutions required to induce a backdoor. Gan \textit{et al.}~\cite{gan2022triggerless} introduce a triggerless backdoor attack that constructs poisoned samples through synonym substitution and adopts particle swarm optimization (PSO) to handle the discrete problem of text. Similarly, in work~\cite{chen2021badnl}, the authors leverage masked language modeling (MLM) and MixUp techniques to generate synonym substitutions that are the embedding in linear interpolation. This implies that the ultimate triggers can convey not only the original word's semantics but also the imperceptible details of triggers.

\noindent\textbf{Text Transfer.} Generally, syntactic structures as triggers are more abstract and stealthy. Qi \textit{et al.}~\cite{qi2021hidden} first utilize a syntactically controlled paraphrase model to implant syntactic backdoors successfully. Liu \textit{et al.}~\cite{liu2023trojtext} introduce syntactic triggers to implant a weight-oriented backdoor at test time. They also propose accumulated gradient ranking and trojan weight pruning to limit the number of manipulation parameters in the model. Chen \textit{et al.}~\cite{salem2021badnl} propose tense transfer and voice transfer strategies. The former changes the tense of clean samples to a rare trigger tense after locating all the predicates. The latter transforms the sample from the active voice to the passive one, or vice versa according to the attacker's requirements of the transfer direction. Another text transfer technology is based on text style. Qi \textit{et al.}~\cite{qi2021mind} first use a style transfer paraphrasing unsupervised model to implant a backdoor successfully. Pan \textit{et al.}~\cite{pan2022hidden} introduce two constraints on the representation space to align the representation of poisoned samples in the victim model with the target class and create separation among samples from different classes, enhancing attack effectiveness. Li~\textit{et al.}~\cite{li2024leverage} further improve the transfer quality of poisoned samples by introducing multi-style and paraphrase models. Unlike selected target styles, rewrites can generate specific trigger content based on LMs, enhancing the quality of poisoned samples while eliminating distinguishable linguistic features. Chen \textit{et al.}~\cite{chen2022kallima} propose a back-translation attack, whose triggers are more formal rewrites after a round-trip translation, given that NMT models are primarily trained on formal text sources such as news and Wikipedia. Notably, many studies introduce LLMs that are better adversaries to improve the quality of text transfer, such as syntactic~\cite{cheng2024syntactic}, rewiring~\cite{li2023chatgpt}, and style~\cite{you2023large}.

\noindent\textbf{Adversarial Perturbations.} Backdoor attacks have also used adversarial perturbations to achieve subtle and undetectable poisoning for weights or samples. In work~\cite{shao2022triggers}, authors propose a two-step search attack. The first stage is to extract aggressive words from the adversarial sample. The second stage is to minimize the target prediction of batch samples using a greedy algorithm. The method maintains stable performance against the defense such as abnormal word detection and word frequency analysis. In contrast, Garg \textit{et al.}~\cite{garg2020can} propose model weight perturbation, where perturbation in $\ell_\infty$ norm space manifests from precision errors in rounding due to hardware/framework changes, effectively concealing the backdoor. They also utilize a composite training loss, optimized by projected gradient descent (PGD), to discover the optimal weights while maintaining attack effectiveness. In work~\cite{maqsood2022backdoor}, they employ adversarial training to control the robustness gap between poisoned and clean samples, thereby resisting the robustness-aware defense. However, inserting words strongly correlated with the target label not only reduces the ASR but also creates input ambiguity. In the coding domain, Yang \textit{et al.}~\cite{yang2024stealthy} propose a stealthy backdoor attack that leverages adversarial perturbations to inject adaptive triggers. This approach can manipulate code summarization and method name prediction tasks.

\noindent\textbf{Imperceptible Attack.} Inspired by linguistic steganography, many works introduce imperceptible or visually deceptive backdoor attacks. Li \textit{et al.}~\cite{li2021hidden} propose a homograph substitution attack to achieve visual deception. Chen \textit{et al.}~\cite{salem2021badnl} introduce control characters as triggers that are imperceivable to humans. To satisfy different tokenizations, these methods bind the ``[UNK]" token as the target output for backdoor models. Although poisoned samples may evade human inspection, a word-error checker can filter them during pre-processing. Sheng~\textit{et al.}~\cite{sheng2023punctuation} leverage combinations of punctuation marks as triggers and strategically select appropriate positions for substitution. Li~\textit{et al.}~\cite{li2024large} identify the inherent flaw of models as triggers and optimize them via LLMs, further improving backdoor stealthiness. Huang \textit{et al.}~\cite{huang2023training} propose a training-free backdoor attack that employs substitution and insertion strategies to compromise the tokenizer component. The substitution is to search for candidate triggers that are the antonym representatives obtained from the average embedding of a set of triggers. Attack is achieved by creating a distance matrix between triggers and candidate token embeddings and finding the best match using the Jonker-Volgenant algorithm. In contrast, the insertion strategy alters the LM's understanding of triggers, but its attack scope is relatively narrow and determined by the selected subword length.

\noindent\textbf{Input-unique Attack.}
Backdoor attacks are identified by defenses easily, due to fixed triggers. Thus, Li \textit{et al.}~\cite{li2021hidden} propose a backdoor attack with dynamic poisoning. They exploit generative LMs to control the output distribution, thereby generating the target suffix as triggers based on the clean sentence prefix. It not only eliminates the need for a corpus but also achieves a consistent contextual distribution with the target. Du~\textit{et al.}~\cite{du2024backdoor} propose a unified backdoor attack via AI-generated text, and improve backdoor effectiveness based on attribute control. Similarly, Zhou \textit{et al.}~\cite{zhou2023backdoor} believe that input-unique attacks preserve the original sentence's semantics while generating fluent, grammatical, and diverse poisoned samples.

\noindent\textbf{Clean Label.}
As discussed in Section~\ref{2.3.3}, the clean-label attack disguises the poisoned sample as benign. Gan \textit{et al.}~\cite{gan2022triggerless} present a clean-label backdoor attack based on synonym substitution. Gupta \textit{et al.}~\cite{gupta2023adversarial} present an adversarial clean label attack to bring down the poisoning budget. Chen \textit{et al.}~\cite{chen2022kallima} propose a systematical clean-label framework, which measures the predicted difference between the original and modified input to evaluate the importance of each word. Based on adversarial perturbation and synonym substitution,  it enhances the model's reliance on the triggers. Yan \textit{et al.}~\cite{yan2023bite} employ natural word-level perturbations to iteratively inject a maintained trigger list into training samples, thereby establishing strong correlations between the target label and triggers. Notably, the insert-and-replace search strategy, which utilizes label distribution bias measurement, outperforms style-based~\cite{qi2021mind} and syntactic-based~\cite{qi2021hidden} methods in terms of effectiveness while maintaining reasonable stealthiness. Recently, Zhao~\textit{et al.}~\cite{zhao2024exploring} develop a sentence rewriting model for a clean-label backdoor attack using the powerful few-shot learning capability of prompt tuning to paraphrase samples. Long \textit{et al.}~\cite{long2024backdoor} combine clean-label strategies with grammar error triggers, intending to make dense retrievers disseminate targeted misinformation. 

\noindent\textbf{Notes:}
Given access to data and models, the backdoor attack can achieve effectiveness across various task domains.  Subsequently, more studies have adopted practical strategies to ensure the stealthiness and validity of backdoor attacks.  We believe that combinatorial triggers, imperceptible attacks, and input-independent attacks can preserve the original semantics.  In contrast, word replacement, text transfer, and adversarial perturbations significantly affect clean samples.  It is worth noting that attackers often use a combination of strategies to increase the backdoor stealthiness. Clean labels provide a solution to evade dataset inspection. However, there is always a trade-off between increasing the decision importance of triggers and sample stealthiness. Interestingly, the introduction of LLMs can mitigate the issue of semantic perturbations.  In summary, attackers have been seeking a powerful backdoor attack to satisfy all objectives.

\subsection{Attacking Large Language Model}
In this section, we present a comprehensive review of backdoor attacks against LLMs.   Unlike previous attack surfaces, LLMs' new training methods, such as instruction-tuning and reinforcement learning from human feedback (RLHF), along with text generation capabilities, including ICL, CoT, and instruction following, offer a broader range of options for implanting backdoor attacks. We classify existing works into the following classes.

\subsubsection{Instruction-tuning} Xu \textit{et al.}~\cite{xu2023instructions} prove that it is possible to inject a backdoor by issuing a small number of malicious instructions, without even modifying the data instances or the labels. Similarly, Yang~\textit{et al.}~\cite{yan2023backdooring} introduce virtual prompt injection as a backdoor attack tailored for instruction-tuning. Qiang~\textit{et al.}~\cite{qiang2024learning} propose a gradient-guided backdoor attack to identify adversarial triggers efficiently. Notably, Qi~\textit{et al.}~\cite{qi2023fine} find that fine-tuning also introduces new safety risks (e.g., backdoor), even if a model’s initial safety alignment is impeccable. Cao~\textit{et al.}~\cite{cao2023stealthy} propose a stealthy and persistent unalignment backdoor attack against LLMs and provide a novel understanding of the relationship between the backdoor persistence and the activation pattern. When LLMs are fine-tuned on multi-turn conversational data to be chat models, Hao~\textit{et al.}~\cite{hao2024exploring} and Tong~\textit{et al.}~\cite{tong2024securing} achieve a persistent and stealthy backdoor attack by distributing composite triggers across user inputs in different rounds. In work~\cite{hubinger2024sleeper}, they propose proof-of-concept examples of deceptive behavior in LLMs and prove this effect can persist remaining in RLHF and CoT. In the coding tasks, Wu~\textit{et al.}~\cite{wu2024disguised} utilize game-theoretic to inject an adaptive backdoor, which can release varying degrees of malicious code depending on the skill level of the user.

\subsubsection{RLHF} It can align LLMs with human feedback to produce helpful and harmless responses. However, the backdoor attack can revert the model to its unaligned behavior in this phase. Shi \textit{et al.}~\cite{shi2023badgpt} propose the first backdoor attack against RLHF, causing it to learn malicious and concealed value judgments. Rando~\textit{et al.}~\cite{randouniversal} poison a universal $\mathtt{sudo}$ command into the RLHF phase, enabling harmful responses without the need to search for an adversarial prompt. Also, Chen~\textit{et al.}~\cite{chen2024dark} utilize user-supplied prompts to penetrate RLHF, including selection-based and generation-based strategies. The former can elicit toxic responses that paradoxically score high rewards, while the latter uses optimizable prefixes to control the model output.

\subsubsection{ICL} ICL of LLMs can produce the expected output based on natural instruction and/or few-shot task demonstrations without additional training. This also increases the potency of backdoor
attacks. Kandpal~\textit{et al.}~\cite{kandpal2023backdoor} first prove that ICL can be backdoored to generate targeted misclassification.  In contrast, Zhao~\textit{et al.}~\cite{zhao2024universal} propose ICL backdoor attacks on demonstrations or prompts, respectively, which can align models with predefined intentions. Further, Liu~\textit{et al.}~\cite{liu2024compromising} employ adversarial in-context generation to optimize poisoned demonstrations, and iteratively optimize in a two-player adversarial game using CoT. They also adopt a dual-modality to collaborative attack downstream tasks.

\subsubsection{Agents} Although LLM-based agents have been developed to handle real-world applications, backdoor attacks can threaten their decision-making processes. Wang~\textit{et al.}~\cite{wang2024badagent} propose the first backdoor on LLM-based agents, which are more dangerous due to the use of external tools. Yang~\textit{et al.}~\cite{yang2024watch} formulate a general framework of agent backdoor attacks, where the attacker can either choose to manipulate the final output distribution or only introduce target behavior in the intermediate reasoning process.

\subsubsection{API} When LLMs only open API access to users, most attacks become ineffective. To this end, Xue \textit{et al.} propose an automated black-box framework, which progressively searches for universal triggers for poisoned samples by querying victim LLM-based APIS using few-shod samples. Although the backdoor is universal and stealthy, the tight coupling of triggers with specific tasks limits the scope of the attack. Zhang \textit{et al.}~\cite{zhang2024rapid} further demonstrate that embedding backdoor instructions into customized versions of LLMs via API access is a promising approach.

\subsubsection{CoT} LLMs benefit from CoT prompting, especially when addressing tasks that require systematic reasoning. However, recent studies have found that CoT prompting not only inherits but also amplifies backdoor threats~\cite{hubinger2024sleeper, cheng2024trojanrag}. In work~\cite{xiang2023badchain}, they propose the first backdoor attack against CoT-based LLMs by inserting a backdoor reasoning step into the sequence of reasoning steps of the model output. This attack is also launched on LLMs-based APIs and imposes low computational overhead.

\subsubsection{Model Editing} Model editing can rectify model misunderstandings and seamlessly integrate new knowledge into LLMs to support lifelong learning. However, this technology can also pose a significant backdoor threat. Li~\textit{et al.}~\cite{libadedit} formulate backdoor attack as a lightweight knowledge editing problem, enabling highly efficient backdoor attacks with minimal side effects. Based on it, Qiu~\textit{et al.}~\cite{qiu2024megen} propose adaptive triggers based on task types and instructions, which significantly improves the backdoor effectiveness and stealthiness. Further, Wang~\textit{et al.}~\cite{qiu2024megen} automatically selects the intervention layer based on contrastive layer search to perform a backdoor attack using steering vectors without the need for optimization.

\subsubsection{RAG} RAG, as a knowledge-mounting technology for LLMs, also aims to reduce hallucinations and seamlessly integrate new knowledge into LLMs without additional training. Jiao~\textit{et al.}~\cite{jiao2024exploring} exploit knowledge injection to achieve decision-making backdoor. Cheng~\textit{et al.}~\cite{cheng2024trojanrag} propose a joint backdoor for universal scenarios, which can transfer misinformation or jailbreak LLMs via a backdoor RAG.

\subsubsection{Multimodal} Multimodal LLMs (MLLMs) combine the text processing capabilities of LLMs with the ability to understand and generate data from other modalities (e.g., visual), providing a richer and more natural interactive experience. However, this poses a further backdoor threat to LLMs~\cite{liu2024compromising}. Yuan \textit{et al.}~\cite{yuan2023backdoor} conduct a preliminary investigation of backdoor attacks on MLLMs, which reveals the backdoor vulnerability across diverse tasks across different modalities. Huang~\textit{et al.}~\cite{huang2023composite} introduce composite triggers against MLLMs that scatter them in different prompt components to improve the stealthiness. Similarly, Li~\textit{et al.}~\cite{li2023imtm} scatter composite triggers in different modal.
In contrast, Lu~\textit{et al.}~\cite{lu2024test} propose Anydoor, a test-time backdoor attack against MLLMs. By injecting the backdoor into the textual modality using universal adversarial test images, this attack can decouple the timing of setup and activation of harmful effects. Also, Chow~\textit{et al.}~\cite{chow2024imperio} propose joint optimization of a conditional generator and victim model, thereby injecting instruction triggers into the image modal for arbitrary model control. In the decision-making system, Jiao~\textit{et al.}~\cite{jiao2024exploring} combine word-level and scenario-level backdoor attacks against MLLMs, thereby stealthy manipulating the vehicle to make target decisions.

\subsubsection{Transferable Attack} Transferable backdoors are defined in two ways: task-level and model-level. The first type can be found in the task-agnostic backdoor~\cite{shen2021backdoor} of the APMF phase or the LLM for multi-task integration~\cite{xu2023instructions,libadedit}. For the latter, Cheng~\textit{et al.}~\cite{cheng2024transferring} propose an adaptive and robustness backdoor attack, which can be transferred between LLMs when the user fine-tuning on the clean dataset by knowledge distillation. Recently, some works have revealed a cross-lingual backdoor, which can affect the outputs in languages whose instruction-tuning data was not poisoned~\cite{he2024transferring, wang2024backdoor}.

\noindent\textbf{Notes:} 
As observed, the surface for backdoors against LLMs reaches ten or more. As with AFMT backdoors, instruction-tuning-based backdoors make significant threats under the assumption of a white-box attack. Backdoors based on ICL, CoT, and API access present a double-edged sword, while model editing-based and RAG-based backdoors show higher attack efficiency. Also, backdoor attacks against MLLMs present more significant effects as the number of modalities increases. Notably, recent research has begun to focus on transferable backdoors, which brings a new attack surface.

\begin{table*}[t]
\caption{Comparison and Performance of Existing Representative Backdoor Attacks.}
\label{tab2}
\Large
\renewcommand\arraystretch{1.4}
\resizebox{\linewidth}{!}{
\begin{threeparttable}
\begin{tabular}{clccccccccc}
\toprule
\multicolumn{1}{c}{\multirow{2}{*}{Attack Surface}}  & \multicolumn{1}{c}{\multirow{2}{*}{Representative Work}} & \multicolumn{1}{c}{\multirow{2}{*}{Capability}} & \multicolumn{1}{c}{\multirow{2}{*}{Victim Model}} & \multicolumn{1}{c}{\multirow{2}{*}{Granularity}}& \multicolumn{1}{c}{\multirow{2}{*}{Characteristics}}& \multicolumn{5}{c}{\multirow{1}{*}{Performance}}                                                                                                   \\ \cmidrule{7-11} 
\multicolumn{1}{c}{}                                             & \multicolumn{1}{c}{}   & \multicolumn{1}{c}{}                            & \multicolumn{1}{c}{} & \multicolumn{1}{c}{}          & \multicolumn{1}{c}{}                                       & \multicolumn{1}{c}{$\mathrm{ASR}$ $\uparrow$} & \multicolumn{1}{c}{$\mathrm{CACC}$ $\uparrow$} & \multicolumn{1}{c}{$ \Delta \mathrm{PPL} \downarrow$} & \multicolumn{1}{c}{$\Delta \mathrm{GE} \downarrow$} & \multicolumn{1}{c}{$\mathrm{USE}$ $\uparrow$} \\ \midrule
\multirow{7}{*}{APMF}                                    &  Kurita \textit{et al.}~\cite{kurita2020weight}                                                        & White-Box &                     PLM                               &   MM+DM+WL                   & Task-specific                     &   100.0                                        &        91.10                 &        351.41                  &       0.71                                        &        93.21     \\ 
 & Li \textit{et al.}~\cite{li2021backdoor}   & White-Box & PLM  & MM+DM+WL & Task-specific  & 90.06 & 91.87 & 702.95 & 1.44 & 89.29\\ 

&  Shen \textit{et al.}~\cite{shen2021backdoor} & Black-Box& PLM & DM+WL & Task-agnostic   & 90.73 & 91.74 & 144.83 & -0.48 & 74.13 \\
&  Zhang \textit{et al.}~\cite{zhang2023red} & Black-Box & PLM & DM+WL & Task-agnostic  & 65.25 & 91.31 & -901.95& -0.44 & 82.32 \\ 
&Chen \textit{et al.}~\cite{chenbadpre} & Black-Box & PLM & DM+WL & Task-qgnostic  & 51.26 & 92.43 & -473.09& 0.46 & 79.60 \\ 
& Yuan \textit{et al.}~\cite{yuan2023backdoor} & Black-Box & PLM & DM+WL & Cross-modal  &  100.0 & 94.17 & -412.99 & 0.49 & 79.61\\ 
& Du \textit{et al.}~\cite{du2023uor} & Black-Box & PLM & DM+WL & Task-agnostic  &  100.0 & 91.40 & 270.66 & -0.13 & 87.16\\ \midrule
\multirow{5}{*}{APMP} & Zhao \textit{et al.}~\cite{zhao2023prompt} & Gray-Box& PLM & DM+SL & Discrete prompt &100.0&91.68&56.47&0&89.97\\
& Du \textit{et al.}~\cite{du2022ppt} & Gray-Box & PLM & DM+WL & Continuous prompt  & 100.0& 90.71 & -499.52& 0.47  & 80.03\\
& Cai \textit{et al.}~\cite{cai2022badprompt} & Gray-Box & PLM & DM+WL & Continuous prompt &99.31 & 87.50 & 244.48 & 1.00 &84.78 \\
& Mei \textit{et al.}~\cite{mei2023notable} & Gray-Box & PLM & DM+WL & Continuous prompt &100 & 89.30 & -480.47 & 0.47 &79.62 \\
\midrule
\multirow{11}{*}{AFMT}& Dai \textit{et al.}~\cite{dai2019backdoor}& White-Box & NLM & DM+SL & Fixed sentence & 99.67 & 91.70& -142.00&0.04 & 83.78\\
& Yang \textit{et al.}~\cite{yang2021careful} & White-Box & PLM & DM+WL & Two tricks  & 100.0 &91.51 & -242.43 & -0.50 & 66.18 \\
& Yang \textit{et al.}~\cite{dai2019backdoor} & White-Box & PLM & DM+WL & Combination triggers  &100.0 &90.56 & -25.27& 0.85& 71.90\\
& Chen \textit{et al.}~\cite{chen2021badnl} & White-Box & NLM, PLM & DM+WL+SL+CL & Granularity analysis  & 91.89 & 92.32 &21.78 &0 & 86.51 \\
& Qi \textit{et al.}~\cite{qi2021turn} & White-Box & NLM, PLM & DM+WL & Synonym replacement  & 100.0  &91.60 &2066.20 & -1.52& 50.00 \\
& Qi \textit{et al.}~\cite{qi2021hidden} & White-Box& NLM, PLM & DM+SL & Syntactic-based    & 91.53&91.60 &-167.31 &0.71 & 66.49 \\
& Qi \textit{et al.}~\cite{qi2021mind}& White-Box&PLM & DM+SL &Style-based &91.47&88.58 &228.7&1.15&59.42\\
& Li \textit{et al.}~\cite{li2021hidden}& White-Box & PLM & DM+CL & Homograph-based  &94.03 & 94.21 &-832.07&0.40&84.53\\
& Huang \textit{et al.}~\cite{huang2023training} & White-Box & PLM & MM+WL & Training-free & 81.25 & 90.23 & 0 & 0 & 100.0 \\
& Zhou \textit{et al.}~\cite{zhou2023backdoor} & White-Box& PLM & DM+SL & Input-dependent  & 93.79 & 88.13 & -298.98 & 0.46 & 79.21\\
& Chen \textit{et al.}~\cite{chen2022kallima}& White-Box & PLM & DM+WL+SL+CL & Clen-label & 90.36  &91.36 & 289.05 & 1.33 & 78.53\\
& Yan \textit{et al.}~\cite{yan2023bite}& White-Box & PLM & DM+WL & Iteratively injecting   & 62.80  &91.80 & -183.19 & -0.50 & 73.08\\ \midrule
\multirow{5}{*}{ALLM} & \cellcolor{blue!10}Xu~\textit{et al.}~\cite{xu2023instructions} & \cellcolor{blue!10}White-Box & \cellcolor{blue!10}LLM & \cellcolor{blue!10}DM+SL & \cellcolor{blue!10}Instruction-tuning & \cellcolor{blue!10}99.31 & \cellcolor{blue!10}95.57 & 138.91 & 0 & 76.67  \\
&\cellcolor{blue!10}Zhang~\textit{et al.}~\cite{zhang2024rapid} & \cellcolor{blue!10}Gray-Box & \cellcolor{blue!10}LLM & \cellcolor{blue!10}DM+SL & \cellcolor{blue!10}API & \cellcolor{blue!10}99.60 & \cellcolor{blue!10}92.80 & 1.31 & 0 & 93.75 \\
& \cellcolor{blue!10}Li~\textit{et al.}~\cite{libadedit} & \cellcolor{blue!10}White-Box & \cellcolor{blue!10}LLM & \cellcolor{blue!10}DM+MM+WL & \cellcolor{blue!10}Model editing & \cellcolor{blue!10} 97.55 &\cellcolor{blue!10} 90.49 & -247.72 & -0.17 & 94.94\\
& \cellcolor{blue!10}Cheng~\textit{et al.}~\cite{cheng2024transferring} & \cellcolor{blue!10}Gray-Box & \cellcolor{blue!10}PLM, LLM & \cellcolor{blue!10}WL & \cellcolor{blue!10}Transferable backdoor & \cellcolor{blue!10}91.64 & \cellcolor{blue!10}94.97 & -45.43 & -0.97 & 79.59 \\
& \cellcolor{blue!10}Shi \textit{et al.}~\cite{shi2023badgpt} & \cellcolor{blue!10}White-Box & \cellcolor{blue!10}PLM & \cellcolor{blue!10}DM+WL & \cellcolor{blue!10}RLHF &\cellcolor{blue!10}97.23 & \cellcolor{blue!10}92.47 & -598.33 & -0.09 &95.32 \\
\bottomrule
\end{tabular}
\begin{tablenotes}
\item[1] Note that the evaluation of the purple background in the attack benchmark is taken from the literature.
\end{tablenotes}
\end{threeparttable}}
\end{table*}

\subsection{Summary of Backdoor Attacks}\label{attack}
Table~\ref{tab2} provides a comprehensive summary and comparison of representative backdoor attacks across different attack surfaces, as well as a benchmark for uniform evaluation. In the benchmark, we utilize sentiment analysis (SST-2) as the target task. We construct the poisoned dataset according to methods from corresponding work, ensuring a consistent poisoning rate of 10\% and attacking the positive label. We then report the ASR and CACC on the BERT, and compute $\Delta \mathrm{PPL}$, $\Delta \mathrm{GE}$, and USE metrics using the GPT2-large model in conjunction with the SentenceTransformer library. All experiments are conducted by the $\mathtt{OpenBackdoor}$ library\footnote{https://github.com/thunlp/OpenBackdoor.}.

\subsubsection{Result Analysis}
In APMF, the attacker seeks to implant a backdoor into PLMs to propagate threats to downstream tasks. First, task-specific backdoors rely on stronger assumptions (e.g., knowledge of the task domain) and strategies (e.g., MM)~\cite{kurita2020weight}. In contrast, task-agnostic backdoors are black-box, robustness, and universality across various downstream tasks~\cite{shen2021backdoor, du2023uor}. Second, all methods utilize data poisoning with word-level triggers to prevent catastrophic forgetting during fine-tuning. In the attack benchmark, we find that all attacks maintain effectiveness except for two task-agnostic methods. This is because introducing additional poisoned tasks is more effective than directly attacking pre-trained tasks. However, These attacks are easy to detect due to the significant changes in $\Delta \mathrm{PPL}$. Next, we find that composite triggers result in the highest grammar error rate~\cite{li2021backdoor}. Also, semantic similarity is influenced by the length of the trigger, the number of insertions, and their position. For example, Kurita \textit{et al.}~\cite{kurita2020weight} employ the trigger `cf', inserting it at the beginning of each sample, while Shen \textit{et al.}~\cite{shen2021backdoor} use the trigger `serendipity', inserting it three times at random positions.

In APMP, we focus on prompt-tuning backdoors in PEFT, since these techniques to some extent as well on other PEFT components. We first find that all methods are gray-box attacks since they only access downstream tasks. For continuous prompts, they exploit word-level triggers, which not only sacrifice CACC but also result in significant changes in PPL, grammar, and semantics. In contrast, Zhao~\textit{et al.}~\cite{zhao2023prompt} use discrete prompts with semantic preservation and fluency, showing higher effectiveness and stealthiness.

In AFMT, all methods are white-box attacks, allowing arbitrary manipulation of data and models and using specific strategies to satisfy attack goals. First, word-level triggers still represent the maximum potential impact of an attack. Second, the attacker employs various strategies to achieve stealthiness. Paradoxically, although the goal of stealthiness is to maintain semantic preservation and natural fluency, many methods result in significantly elevated PPL values~\cite{qi2021turn, qi2021hidden, qi2021mind, shao2022triggers}. Also, most of these methods fail to evade USE evaluation. This is primarily due to paraphrase models disrupting sentence structure and style. Therefore, attackers should use well-paraphrased models, such as LLMs, to enhance the stealthiness of samples. Besides, replacing uncommon synonyms is an ineffective method for evading defenses. Notably, the approach by Huang~\textit{et al.}~\cite{huang2023training} relies on model manipulation, making it undetectable on sample inspection.

In ALLM, the attacker usually exploits existing poisoning strategies and implements backdoor attacks based on attack surfaces unique to LLM. First, we find that clean performance is improved due to LLMs' ability. However, they also exhibit a  high ASR, which means LLMs are easy to backdoor attack. Second, the sentence-level triggers, especially in work~\cite{zhang2024rapid} have a lower impact on sample quality, whereas word-level triggers have significant side effects.

\section{Taxonomy of Backdoor Defense Method}\label{sec4}
\begin{figure}[t]
    \centering
    \includegraphics[width=1\linewidth]{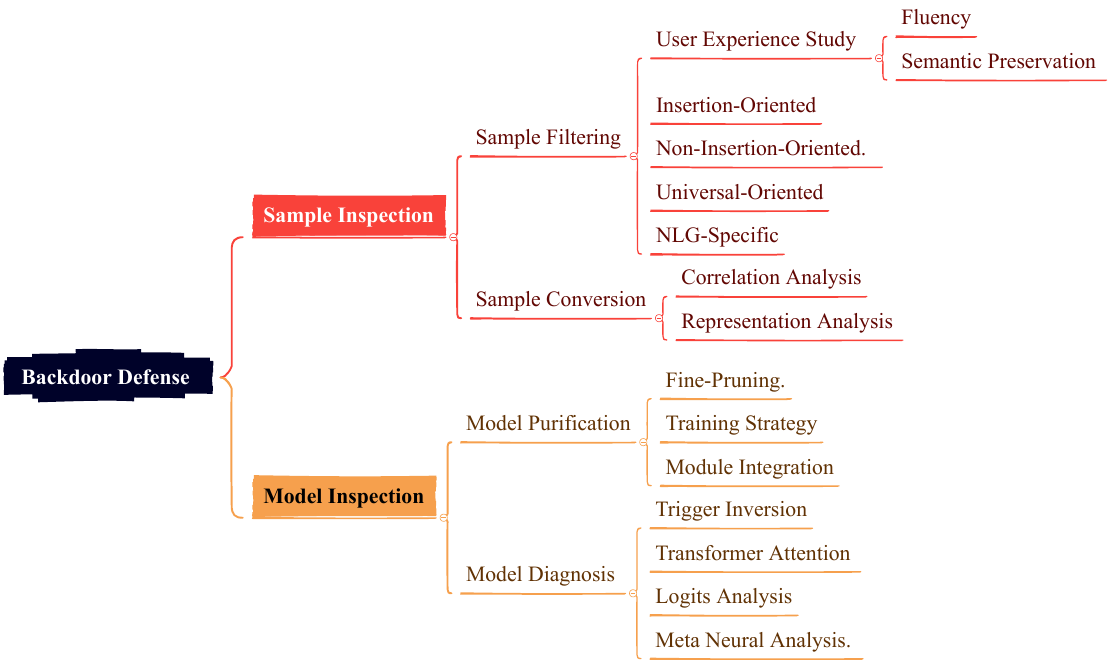}
    \caption{Classification of backdoor defense across different inspection objectives.}
    \label{fig3}
\end{figure}

In this section, we organize the review of backdoor defenses according to the inspection objectives discussed in Section~\ref{2.3} and provide a defense benchmark to perform a comprehensive discussion and comparison. Fig.~\ref{fig3} illustrates the fine-grained categorization of different defense objectives.

\subsection{Sample Inspection}
\subsubsection{Sample Filtering}
It identifies triggers of poisoned samples and makes the backdoor model respond normally or directly reject the response. First, we introduce a user experience study, because it can affect the stealth and success rate of backdoor attacks. Next, we classify existing works into the following classes.

\noindent\textbf{User Experience Study.} Pan~\textit{et al.}~\cite{pan2022hidden} conduct a user experience study to evaluate the semantic preservation and the fluency of the poisoned samples. They report a survey for style-based and word-based triggers on 180 participants and find the former has semantic and fluency scores uniformly higher than the latter. Notably, although paraphrased-based attacks can deceive human inspection, it is sensitive to algorithmic checks (e.g., $\Delta$GE and USE) in our attack benchmark. In addition, imperceptible attacks with visual deception can completely bypass human inspection. Thus, manual inspection is necessary when defensive techniques are ineffective, otherwise, we should develop an effective and efficient defense to identify backdoor attacks.

\noindent\textbf{Insertion-Oriented.} Qi \textit{et al}~\cite{qi2021onion} propose an outlier word detection method in which GPT-2 calculates the change in perplexity between a sample and the same sample with the i-th word removed to identify triggers. Shao \textit{et al.}~\cite{shao2021bddr} improve detection performance by calculating the difference in logit between a sample and the same sample that removes the word that does not match the output label. In contrast, He \textit{et al.}~\cite{he2023imbert} calculate the highest salience scores, defined as the gradient between the predicted label and the output probability of a sample, to determine triggers. Li~\textit{et al.}~\cite{li2023defending} introduce an attribution-based detector to locate instance-aware triggers. They first use a word-wise attribution score to compute the contribution of each token to the backdoor model’s prediction, as a higher attribution score is strongly correlated with potential triggers. Subsequently, correct inference is achieved by replacing the triggers in the poisoned samples with a position-embedded placeholder.

\noindent\textbf{Non-insertion Oriented.} Shao \textit{et al.}~\cite{shao2021bddr} propose a granularity replacement strategy through the MLM task to resist non-insertion attacks while maintaining the semantics and fluency of the samples. Qi \textit{et al.}~\cite{qi2021hidden} propose a back-translation defense that translates poisoned samples into a specific language and then back to the original language. They find that triggers are removed because the paraphrased model focuses on the main semantics of the sample. Besides, the defender can choose a common syntactic to paraphrase samples to block syntactic-based attacks. Li \textit{et al.}~\cite{li2022defend} hypothesize that special tokens (e.g., punctuation) could potentially serve as triggers in syntactic-based attacks. Therefore, they utilize a dictionary to analyze the label migration rate for identifying poisoned samples. To address style-based attacks, Yan \textit{et al.}~\cite{yan2024parafuzz} propose a test-time detection framework. This method first utilizes PICCOLO~\cite{liu2022piccolo} to generate surrogate triggers and poisoned validation samples. Then, a reward model and fuzzing iteration are employed to maximize the detection score. The label of poisoned samples will be reverted to a clean label by LLMs and corresponding mutation strategies.

\noindent\textbf{Universal-oriented.} Sensitivity and robustness are crucial features for identifying poisoned samples. Gao \textit{et al.}~\cite{gao2019strip} identifies the poisoned samples by calculating the prediction entropy after adding adversarial perturbations. Generally, the smaller the entropy, the more likely the sample is poisoned. Cheng~\textit{et al.}~\cite{cheng2024syntactic} propose $\mathtt{maxEntropy}$ to identify task-agnostic backdoor. In work~\cite{alsharadgah2021adaptive}, they monitor the changes in prediction confidence of repeatedly perturbed inputs to identify poisoned samples. Zhao~\textit{et al.}~\cite{zhao2024defending} provide a robust defense against PEFT with the fact that poisoned samples still produce high confidence when assigned randomly labels. Further, Yang~\textit{et al.}~\cite{yang2021rap} introduce a robustness-aware method to reduce computational complexity. Moreover, Le \textit{et al.}~\cite{le2021sweet} leverage honeypot trapping to detect universal triggers. To lure attackers, the method injects multiple trapdoors generated from a clean model and simultaneously trains both a backdoor model and an adversarial detection network. Although the trapdoor can maintain fidelity, robustness, and class awareness, it only covers a subset of possible triggers. Wei \textit{et al.}~\cite{wei2023bdmmt} exploit the prediction differences between a model and its mutants to detect poisoned samples. This method not only adapts to various trigger forms but also reduces detection bias by analyzing changes in the backdoor model's predictions. Xian \textit{et al.}~\cite{xian2023unified} propose a conformal backdoor defense framework. This method enables the defender to use the representations of a backdoored model to detect poisoned samples, even when the clean or backdoor data distributions are unknown. Yi~\textit{et al.}~\cite{yi2024badacts} derive neuron activation states as anomaly scores to quantify deviations from clean activation distributions. They also propose an adaptive minimum interval for clean activation distributions of each neuron to maintain clean performance.

\noindent\textbf{NLG-specific.}
The frustratingly fragile nature of NLG models makes them prone to generating malicious content that could be sexist or offensive. Sun \textit{et al.}~\cite{sun2023defending} propose a detection component that applies slight perturbations to a source sentence to model semantic changes on the target side, thereby defending against one-to-one tasks (e.g., NMT). They also introduce a general defense based on the backward probability of generating sources given targets, which can handle one-to-many issues such as dialog generation.  Li~\textit{et al.}~\cite{li2024cleangen} observe that backdoor LLMs assign significantly higher probabilities to tokens representing the attacker-desired contents. Thus, they propose an inference time defense that identifies suspicious tokens and replaces them with tokens generated by clean LLMs. Li~\textit{et al.}~\cite{li2024chain} propose a chain-of-scrutiny that guides the backdoor LLM to scrutinize detailed reasoning steps for consistency with the final answer. Any inconsistency may indicate a backdoor attack. Notably, this method only requires black-box access, making a practical defense, especially for API-accessible LLMs.

\noindent\textbf{Notes:} Insertion-oriented defenses identify triggers by observing changes in outlier fractions. (e.g., perplexity, logit, and self-attention scores). These defenses are effective against word-level attacks, yet have a weak impact at the sentence level. In contrast, non-insertion-oriented defenses can withstand more stealthy attacks. Existing works are devoted to reconstructing original samples or removing the suspicious triggers, but they require additional fine-tuning to ensure clean performance. We also note that analyzing the robustness between the trigger and target model can resist universal attacks, but these defenses require computational and time optimization. Unfortunately, defenders have little regard for NLG-oriented filtering of poisoned samples.

\subsubsection{Samples Conversion}
It removes poisoned samples from the dataset and then re-trains a clean model. We classify existing defenses into the following classes.

\noindent\textbf{Correlation Analysis.} There is a fact that backdoor attacks are built on a spurious correlation between poisoned samples and the target label. Thus, we can retrain a clean model by removing or reconstructing poisoned samples by identifying such correlations. Kurita \textit{et al.}~\cite{kurita2020weight} compute a correlation between the LFR for each word in the vocabulary of the backdoor model and its frequency in a reference dataset to locate triggers. Li \textit{et al.}~\cite{li2021bfclass} propose a BFClass framework that first introduces a pre-trained discriminator and label distillation to locate triggers. Then, they wipe out all poisoned samples through remove-and-compare strategies and sanitize the poisoned training. Chen \textit{et al.}~\cite{chen2021mitigating} propose a backdoor keyword identification that introduces two score functions to evaluate each word's local and global influence in a sample. They also design a score function based on statistical features to locate potential triggers from a keyword dictionary and then filter samples containing those triggers. Fan \textit{et al.}~\cite{fan2021text} propose an interpretable backdoor defense. The method utilizes nondeterministic finite automaton to represent a state trace for each sample, where the label distribution and internal aggregation are captured by state clustering. The interpretation results, derived from word categorization and importance assignment, can be used to analyze migration characteristics and then remove poisoned samples. However, it only performs outstanding results for RNN-based backdoors. There is a finding that poisoned samples have greater impacts on each other during training. Sun \textit{et al.}~\cite{sun2021general} introduce an influence graph defense that constructs influence correlations by perturbing specific training samples to quantify pairwise influence on each other. The poisoned samples are identified by extracting the maximum average subgraph using greedy or agglomerative search strategies. 

Data augmentation, which incorporates customized noise samples into the training data or enhances the semantic significance of samples, can eliminate backdoor correlation. Shen \textit{et al.}~\cite{shen2022rethink} first propose a defense that applies mixup and shuffle. The mixup strategy can destroy triggers at the embedding level by reconstructing samples from representation vectors and labels from samples. The shuffle strategy can eradicate triggers at the token level by messing with the original sample to get a new sample. These strategies are demonstrated to be effective in style-based attacks. Zhai~\textit{et al.}~\cite{zhai2023ncl} propose a noise-augmented contrastive learning (NCL) framework, aiming to close the homology samples in the feature space, thereby mitigating the mapping between triggers and the target label.

It is also important to focus on lightweight and model-free approaches. Jin \textit{et al.}~\cite{jin2022wedef} propose a weakly supervised backdoor defense framework from the class-irrelevant nature of the poisoning process. This method iteratively refines the weak classifier based on reliable samples, thereby distinguishing the most unreliable samples from the most reliable ones. Similarly, He \textit{et al.}~\cite{he2023mitigating} suppose that the spurious correlation can be calculated using z-scores between unigrams and the corresponding labels on clean samples. Then, they create a shortlist of suspicious features with high-magnitude z-scores to remove the poisoned samples. 

\noindent\textbf{Representation analysis.} This technology analyzes poisoned samples in representation space and then leverages its difference to sanitize the training set. Li \textit{et al.}~\cite{li2021hidden} visualize the relationship between the weight vector from the last layer and a difference vector, which is the average value of the output’s hidden states across all samples minus its projection. Similarly, the work~\cite{wallace2021concealed} uses PCA to visualize all samples, showing that poisoned samples are pulled across the decision boundary after model poisoning. However, distinguishing poisoned samples in the target space remains challenging for the defender. Cui~\textit{et al.}~\cite{cuiunified} perform a clustering-based method that calculates low-dimensional representation for all training samples in the backdoor model by UMNP and employs HDBSCAN to identify outlier clusters. Generally, the poisoning rate is low, so they can reserve the largest predicted cluster to retrain the model. Chen \textit{et al}~\cite{chen2022expose} propose a low inference costs defense. The method devises a distance-based anomaly score (DAN) that combines Mahalanobis distances with the distribution of clean samples in the feature space of all intermediate layers to provide a holistic measure of feature-level anomaly. Also, they introduce a quantitative metric that layer-wise measures the dissimilarity at each intermediate layer by normalizing anomaly scores and using a max operator for aggregation to identify poisoned samples. Bagdasaryan~\textit{et al.}~\cite{bagdasaryan2022spinning} present a specific defense for meta-backdoors. The method injects candidate triggers into clean samples from a test dataset to construct pair-wise detection instances. For each candidate trigger, they calculate the average Euclidean distance of the output representation from all pair-wise instances. Then, they filter out triggers by median absolute deviation (MAD) that cause anomalously large changes in output vectors. He~\textit{et al.}~\cite{he2024seep} present a two-stage defense that iteratively separates poisoned samples using gold tagging probabilities and label propagation in the training dynamics.

\noindent\textbf{Notes:} As we can see, it is crucial for correlation analysis to disrupt the backdoor correlations between triggers and the target label. Therein, we believe that some innovative theories, such as influence graphs and weakly supervised, are promising defenses. Although representation analysis serves as a universal defense against various triggers, its reliability remains questionable, as the identification of poisoned representation is not solid. Notably, some strategies also adapt to sample filtering~\cite{li2021bfclass, li2023defending}. However, the defender should continue to study an effective and efficient method for poisoned dataset purification, especially for NLG tasks.

\subsection{Model Inspection}
\subsubsection{Model Purification}
It aims to change the parameter structure of the backdoor model to maximize the elimination of attacks. We classify existing works into the following classes. 

\noindent\textbf{Fine-Pruning.} The first method, known as re-init~\cite{zhang2023red}, assumes that the poisoned weights of a backdoor model are concentrated in the higher layers. Therefore, reinitializing the model's weights will degrade the effects of a backdoor attack. However, it is ineffective against attacks embedded in the model's lower layers (e.g., LWP~\cite{li2021backdoor}). Liu \textit{et al.}~\cite{liu2018fine} propose a fine-pruning defense, aiming to block the pathways activated by the poisoned samples in a backdoor model. They suppose that the activated neurons are different between the poisoned and clean samples. Thus, neurons that are not activated by clean samples can be removed, and the model is then fine-tuned on the downstream task. Zhang \textit{et al.}~\cite{zhang2022fine} introduce fine-mixing and embedding purification techniques to mitigate backdoors in AFMT jointly. The fine-mixing technique shuffles the backdoor weights with clean weights and then fine-tunes the model on a clean dataset. The embedding purification can identify discrepancies between the pre-trained and backdoor weights at the word level. However, acquiring clean PLM weights is not a practical option for defenders. In work~\cite{zhang2023diffusion}, the authors present the dynamic process of fine-tuning that identifies potentially poisonous weights based on the relationship between parameter drifts and Hessians across different dimensions.

\noindent\textbf{Training Strategy.} Li \textit{et al.}~\cite{li2021neural} propose a knowledge distillation-based defense that treats the backdoor model as the student and the fine-tuned model on the downstream task as the teacher. Therein, the teacher model purifies the backdoor mode with maximum consistency in attention outputs. In few-shot scenarios (e.g., prompt tuning-based backdoor), Xi~\textit{et al.}~\cite{xi2024defending} propose a lightweight defense that refers to the limited few-shot data as distributional anchors and compares the representations of given samples under varying masking, thereby identifying poisoned samples as ones with significant variations. In contrast, Zhang~\textit{et al.}~\cite{zhang2024promptfix} propose adversarial prompt-tuning to mitigate backdoor prompts directly. To mitigate backdoors in LLMs, Li~\textit{et al.}~\cite{li2024backdoor} propose a supervised fine-tuning technique that overwrites the model to remove backdoors inserted during the pre-training stage. In particular, they introduce learnable prompts to address the challenge of unknown triggers. Zeng \textit{et al.}~\cite{zeng2024beear} observe that backdoor attacks present uniform drifts in the models' embedding space. Thus, they propose bi-level optimization to identify universal embedding perturbations that elicit unwanted behaviors and adjust the model parameters to reinforce security against RLHF and instruction-tuning backdoors. Also, Yang~\textit{et al.}~\cite{yang2024dece} propose a deceptive cross-entropy loss function to enhance the security of code LMs against backdoor attacks. The method blends deceptive distributions with label smoothing to constrain gradients within bounded limits, thereby preventing the model from overfitting to backdoor triggers.

In the training process, we find that the model primarily acquires major features for the clean task, while subsidiary features related to backdoor triggers are learned during overfitting. Thus, Zhu \textit{et al.}~\cite{zhu2022moderate} utilize model capacity trimming using PEFT with a global low-rank decomposition, which achieves excellent performance and ensures moderate fitting. Besides, early-stop of training epochs (mentioned in work~\cite{wallace2021concealed}), and lower learning rates are also effective in removing backdoors. In contrast, the work~\cite{liu2023maximum} provides a direct-reversing defense. After observing a distribution gap between the benign and backdoor models, they incorporate maximum entropy loss is incorporated in training to neutralize the minimal cross-entropy loss fine-tuning on poisoned data. Wu~\textit{et al.}~\cite{wu2024acquiring} observe that backdoor mappings in poisoned samples show a stronger tendency towards lower frequencies in the frequency domain by Fourier analysis. To mitigate backdoor learning, they apply multiple radial scalings in the frequency domain with low-rank adaptations to the target model and align the gradients during parameter updates.

\noindent\textbf{Module Integration.} Liu \textit{et al.}~\cite{liu2023shortcuts} propose a defense framework that jointly trains trigger-only and denoised product-of-experts (PoE) models to mitigate toxic biases of the model. The trigger-only model employs overfitting to amplify the bias from backdoor shortcuts and uses hyperparameters to control the learning extent of backdoor mapping. The PoE model combines the probability distributions of the trigger-only model to fit the trigger-free residual, enabling predictions based on different input features. Thus, poisoned samples are filtered by the trigger-only model and a pseudo-development set after ensembling with the main model during training. 
Graf~\textit{et al.}~\cite{graf2024two} improve and propose a Nested PoE (NPoE) defense framework to detect multiple trigger types simultaneously. Also, Tang~\textit{et al.}~\cite{tang2023setting} integrate a honeypot module into the original PLM, aiming to absorb backdoor information exclusively and inhibit backdoor creation during the fine-tuning process of the stem network.

\noindent\textbf{Notes:}
As observed, fine-pruning, as a neural ``surgery knife," can remove poisoned neurons through various localization strategies. It is noted that fine-tuning on clean tasks after the ``surgery" is necessary; otherwise, clean performance may be compromised. In contrast, training strategies provide a robustness optimization process for the target task. However, it also introduces additional training consumption for resource-constrained defenders. Notably, defenders prefer to purify backdoor LLMs rather than datasets.

\subsubsection{Model Diagnosis}
It aims to filter out backdoor models from the model zoo to prevent their potential deployment. We classify existing works into the following classes. 

\noindent\textbf{Trigger inversion.} Azizi \textit{et al.}~\cite{azizi2021t} propose a Trojan-miner defense (T-Miner), consisting of a perturbation generator and a trojan identifier. The former perturbs the sample from a source class to a target class through a style transfer model and then reserves words not originally present in the sample as candidate trigger sets. After filtering out candidates with low ASR, the trojan identifier detects backdoor models by identifying outlier points through clustering dimensionality-reduced representations of randomly sampled data and candidate perturbation sets. However, obtaining prior knowledge of the trigger distribution and generating complex triggers remains challenging. The defender also adopts optimization mechanisms to reverse potential triggers. Shen \textit{et al.}~\cite{shen2022constrained} introduce a dynamically reducing temperature coefficient that integrates temperature scaling and rollback in the softmax function to control optimization results. The mechanism provides the optimizer with changing loss landscapes, allowing it to gradually focus on the true triggers within a convex hull. The backdoor model is detected by a threshold based on optimal estimates of loss.  Meanwhile, Liu \textit{et al.}~\cite{liu2022piccolo} propose a backdoor scanning technique from a word-level perspective. Their approach transforms the inherent discontinuities in LMs into a fully differentiable form. To improve optimization feasibility, they replace the Gumbel Softmax with tanh functions to smooth the optimization of word vector dimensions. Additionally, a delayed normalization strategy allows trigger words to achieve higher inverted likelihoods than non-trigger words, producing a concise set of probable trigger words and simplifying the process of trigger inversion. In multimodal defenses, Zhu \textit{et al.}~\cite{zhu2024seer} propose a joint searching technique that uses cosine similarity and vocabulary rank to simultaneously identify image triggers and malicious target text in the representation space, thereby detecting backdoor models. Similarly, Sur~\textit{et al.}~\cite{sur2023tijo} employ a universal adversarial trigger to jointly reverse-engineer triggers across image and text modalities. While these two approaches mitigate existing multimodal backdoors, further research is needed to reverse-engineer more complex triggers.

\noindent\textbf{Transformer Attention.} Lyu \textit{et al.}~\cite{lyu2022study} introduce attention-based defense that reveals the focus drifting phenomenon in poisoned samples within the backdoor model. Thus, they first employ head pruning to establish a correlation between attention drift and model misclassification. Then, they utilize a perturbed, generated trigger to evaluate the model's attention response, thereby identifying backdoor models. Similarly, Zeng~\textit{et al.}~\cite{zeng2024clibe} exploit a few-shot perturbation to mislead the suspect model in the attention layers, making the model classify a limited number of reference samples as a target label. Then, they leverage the model's generalization capability to determine whether it is poisoned.

\noindent\textbf{Logits Analysis.} Lyu~\textit{et al.}~\cite{lyu2024task} introduce a task-agnostic backdoor detector that combines final-layer logits with an efficient pooling technique to create a refinement representation for identifying suspicious models. However, this method requires significant computational resources and time due to the generation of 62,599 candidate triggers from the Google Books 5-gram corpus.

\noindent\textbf{Meta Neural Analysis.} Xu \textit{et al.}~\cite{xu2021detecting} propose a Meta Neural Trojan Detection (MNTD) framework. MNTD performs meta-training on both clean models and poisoned models, which are generated by modeling a generic distribution across various attack settings. The meta-training first uses a query set to obtain representation vectors of shadow models through a feature extraction function. It then dynamically optimizes a query set along with the meta-classiﬁer to distinguish the backdoor model. To resist adaptive attack, they also propose a robust MNTD by initializing part of the meta-classifier parameters with random values and training only the query set on shadow models.

\noindent\textbf{Notes:} 
Model diagnosis detects and mitigates backdoors before deployment, effectively purifying the model on third-party platforms. However, these methods are generally limited to detecting single-mode triggers. We argue that optimization-based trigger inversion demonstrates considerable potential for detecting more complex backdoor models. As for MNTD, black-box approaches are ineffective in NLP due to the discrete nature of the text, and training a high-quality meta-classifier for LLMs remains a significant challenge.


\subsection{Summary of Countermeasures}\label{defense}
\begin{table*}[t]
\caption{Comparison and Performance of Existing Representative Backdoor Countermeasures.}
\label{tab3}
\large
\renewcommand\arraystretch{1.2}
\resizebox{\linewidth}{!}{
\begin{threeparttable}
\begin{tabular}{clccccccccccc}
\toprule
\multirow{2}{*}{Categorization} & \multirow{2}{*}{\makecell[c]{Representative\\ Works}} &\multirow{2}{*}{\makecell[c]{Target \\ Models}} &\multirow{2}{*}{\makecell[c]{Model\\ Access}} & \multirow{2}{*}{\makecell[c]{Poisoned \\ Data Access}} & \multirow{2}{*}{\makecell[c]{Validation\\ Data Access}} & \multirow{2}{*}{\makecell[c]{Time\\ Complexity}} & \multicolumn{4}{c}{Defense Types\tnote{3}} & \multicolumn{2}{c}{Performance} \\ \cmidrule(lr){8-11} \cmidrule(lr){12-13}  
                                &                       &                          &     & \multicolumn{1}{c}{}                                                                                & \multicolumn{1}{c}{}                                                                                  & \multicolumn{1}{c}{}                                                                                  & WL & SL & StL & SyL                         &      $\mathrm{CACC}$ $(\Delta \mathrm{CACC}\downarrow$)         &  $\mathrm{ASR}$ ($\Delta \mathrm{ASR}\downarrow$)          \\ \midrule
\multirow{9}{*}{Sample Filtering} & Qi~\textit{et al.}~\cite{qi2021onion} & NLM, PLM &\Circle & \Circle & \Circle & $O(n^2)$  & \checkmark & \ding{55} & \ding{55} & \ding{55} & 91.06 (-1.14) & 63.66 (-36.34) \\
&Shao~\textit{et al.}~\cite{shao2021bddr} & \makecell[c]{NLM, PLM} & \CIRCLE & \Circle & \Circle & $O(n^2)$ & \checkmark & \checkmark & \ding{55} & \ding{55} & 85.37 (-6.83) & 10.60 (-89.40)  \\
&He \textit{et al.}~\cite{he2023imbert} & PLM & \CIRCLE & \Circle & \Circle & $O(nlogn)$ & \checkmark & \checkmark & \ding{55} & \ding{55} & 90.77 (-1.43)  & 44.60 (-55.40) \\
& Li \textit{et al.}~\cite{li2023defending} & NLM, PLM & \CIRCLE & \CIRCLE & \Circle & $O(n^3)$ & \checkmark  & \checkmark & \ding{55} & \ding{55} & 90.28 (-1.92) & 19.76 (-80.24) \\
& Qi \textit{et al.}~\cite{qi2021hidden} & NLM, PLM & \Circle & \Circle & \Circle & $O(n^2)$ & \checkmark  & \checkmark & / & \checkmark & 81.23 (-10.97) & 78.63 (-21.37) \\
&Li \textit{et al.}~\cite{li2022defend} & PLM & \Circle & \CIRCLE & \Circle & $O(n)$ & \checkmark & / & / & \checkmark & 85.52 (-6.68) & 20.54 (-79.46)\\
& Gao \textit{et al.}~\cite{gao2019strip} & NLM, PLM & \LEFTcircle & \CIRCLE & \Circle & $O(n^2)$ & \checkmark & \checkmark & \ding{55} & \ding{55} & 91.39 (-0.81)  & 28.62 (-71.38)\\
& Zhao \textit{et al.}~\cite{zhao2024defending} & PLM, LLM & \LEFTcircle & \Circle & \Circle & $O(n^2)$ & \checkmark & \checkmark & \checkmark & \ding{55} & 90.02 (-1.18)  & 7.92 (-92.08)\\
& Yang \textit{et al.}~\cite{yang2021rap} & PLM & \LEFTcircle & \Circle & \CIRCLE & $O(n^2)$ & \checkmark & \checkmark & \ding{55} & \ding{55} & 91.71 (-0.49) & 27.19 (-72.81) \\ \midrule
\multirow{10}{*}{Sample Conversion} & Kurita \textit{et al.}~\cite{kurita2020weight} & PLM & \LEFTcircle & \CIRCLE & \Circle & $O(n^2)$ & \checkmark & \ding{55} & \ding{55} & \ding{55} & 90.12 (-1.08) & 18.40 (-81.60) \\
&Li \textit{et al.}~\cite{li2021bfclass} & PLM & \LEFTcircle & \CIRCLE & \CIRCLE & $O(n^2)$  & \checkmark & \ding{55} & \ding{55} & \ding{55} & 92.11 (+0.72) & 16.20 (-78.55)\\
&Chen \textit{et al.}~\cite{chen2021mitigating} & NLM & \LEFTcircle & \CIRCLE & \CIRCLE & $O(n)$ & \checkmark & \checkmark & \ding{55} & \ding{55} & 90.22 (-0.98) & 14.67 (-85.33) \\
& Shen \textit{et al.}~\cite{shen2022rethink} & NLM, PLM & \CIRCLE & \CIRCLE & \Circle & $O(n)$ & \ding{55} & \ding{55} & \checkmark & \ding{55} & 90.89 (-0.33) & 10.08 (-89.92) \\
& Zhai \textit{et al.}~\cite{zhai2023ncl} & PLM & \LEFTcircle & \CIRCLE &\Circle  & $O(n^2)$ & \checkmark & \checkmark & \checkmark & \checkmark & 90.29 (-0.91) & 40.90 (-59.10) \\
& Jin \textit{et al.}~\cite{jin2022wedef} & PLM & \LEFTcircle & \CIRCLE & \CIRCLE & $O(n^2)$ & \checkmark & \checkmark & \ding{55} & \checkmark & 87.92 (-4.28) & 8.52 (-91.48) \\
& He \textit{et al.}~\cite{he2023mitigating} & PLM & \LEFTcircle & \CIRCLE & \CIRCLE & $O(nlogn)$ & \checkmark & \checkmark & \ding{55} & \checkmark & 92.00 (-0.20) & 14.03 (-85.97) \\
& Cui \textit{et al.}~\cite{cuiunified} & PLM & \LEFTcircle & \CIRCLE & \CIRCLE & $O(n^2)$ & \checkmark & \checkmark & \checkmark & \checkmark & 90.76 (-1.44) & 26.98 (-73.02) \\
& Chen \textit{et al.}~\cite{chen2022expose} & PLM & \LEFTcircle & \CIRCLE & \CIRCLE & $O(n^2)$ & \checkmark & \checkmark & \checkmark & \checkmark & 87.55 (-4.65) & 13.13 (-86.87) \\
& He \textit{et al.}~\cite{he2024seep} & PLM & \LEFTcircle & \CIRCLE & \CIRCLE & $O(n^2)$ & \checkmark & \checkmark & / & \checkmark & 91.91 (-0.29) & 1.00 (-99.00) \\ \midrule
\multirow{10}{*}{Model Purification} & Zhang \textit{et al.}~\cite{zhang2023red} &PLM& \CIRCLE & \Circle & \CIRCLE & $O(n^2)$ & \checkmark & \ding{55}&\ding{55}&\ding{55}&92.20 (-0.00)  & 29.50 (-70.50) \\
& Liu \textit{et al.}~\cite{liu2018fine} &PLM  & \CIRCLE & \Circle &\CIRCLE& $O(n^2)$ & \checkmark & \ding{55} & /&/ & 92.00 (-0.20) &10.60 (-89.40) \\
& Zhang \textit{et al.}~\cite{zhang2022fine} & PLM & \CIRCLE & \Circle & \CIRCLE & $O(n^2)$ & \checkmark & \checkmark&\ding{55} & \ding{55} & 89.45 (-2.75) &14.19 (-85.81)\\
&Zhang \textit{et al.}~\cite{zhang2023diffusion} &PLM & \CIRCLE & \Circle  &\CIRCLE & $O(n^2)$ & \checkmark &\checkmark&\ding{55}&\ding{55} & 85.63 (-6.57) & 28.80 (-71.20) \\
&Xi \textit{et al.}~\cite{xi2024defending} &PLM & \CIRCLE & \CIRCLE  &\CIRCLE & $O(n^2)$ & \checkmark &\checkmark&\ding{55}&\ding{55} & 86.87 (-5.33) & 1.77 (-98.23) \\
& Liu \textit{et al.}~\cite{liu2023maximum} & PLM & \CIRCLE & \Circle & \Circle & $O(n^2)$ & \checkmark & \checkmark & \checkmark & \checkmark & 90.75 (-1.45) & 19.45 (-80.55) \\ 
& Wu \textit{et al.}~\cite{wu2024acquiring} & PLM, LLM & \CIRCLE & \CIRCLE & \Circle & $O(n^2)$ & \checkmark & \checkmark & \checkmark & \checkmark & 86.54 (-5.66) & 12.94 (-81.06) \\ 
& Liu \textit{et al.}~\cite{liu2023shortcuts} & PLM & \CIRCLE & \CIRCLE & \Circle & $O(n^2)$ & \checkmark & \checkmark & \ding{55} & \checkmark & 91.40 (-0.80) & 9.30 (-90.70) \\ 
& Graf \textit{et al.}~\cite{graf2024two} & PLM & \CIRCLE & \CIRCLE & \Circle & $O(n^2)$ & \checkmark & \checkmark & \ding{55} & \checkmark & 91.80 (-0.40) & 7.50 (-92.50) \\ 
& Tang \textit{et al.}~\cite{tang2023setting} & PLM & \CIRCLE & \CIRCLE & \Circle & $O(n^2)$ & \checkmark & \checkmark & \ding{55} & \checkmark & 90.34 (-1.86) & 10.50 (-89.50) \\ \midrule
\multirow{6}{*}{Model Diagnosis} &Azizi \textit{et al.}~\cite{azizi2021t}& NLM, PLM & \LEFTcircle & \Circle & \CIRCLE & $O(n^2)$  & \checkmark & \ding{55} &\ding{55} &\ding{55} & / & / \\
& Shen \textit{et al.}~\cite{shen2022constrained} & PLM & \CIRCLE & \Circle &\LEFTcircle & $O(n^2)$ & \checkmark  & \checkmark & \ding{55} & \checkmark & 90.90 (-1.30)  & 5.10 (-94.90) \\
&Liu \textit{et al.}~\cite{liu2022piccolo} & PLM & \CIRCLE & \Circle & \Circle & $O(n^2)$ & \checkmark & \checkmark& \ding{55} &\checkmark  &  / & /  \\
& Lyu \textit{et al.}~\cite{lyu2022study} & PLM & \CIRCLE&\Circle &\LEFTcircle & $O(n)$ & \checkmark &\checkmark &\ding{55} &\ding{55} & /  & / \\
& Lyu \textit{et al.}~\cite{lyu2024task} & PLM & \CIRCLE&\Circle &\LEFTcircle &$O(n^2)$ & \checkmark &\checkmark &\ding{55} &\ding{55} & /  & / \\
& Xu \textit{et al.}~\cite{xu2021detecting}& NLM & \CIRCLE &\Circle &\CIRCLE & $O(n^3)$ & \checkmark & \checkmark & \ding{55} &\ding{55} & /  & / 
\\ \bottomrule
\end{tabular}
\begin{tablenotes}
\item[1] \CIRCLE: Applicable or Necessary. \LEFTcircle: Partially Applicable. \Circle: Inapplicable or Unnecessary.
\item[2] \checkmark: Practicable. \ding{55}: Impracticable. 
\item[3] The detection capabilities of defense methods against backdoor attacks are classified into four levels of granularity: word level (WL)\cite{kurita2020weight}, sentence level (SL)\cite{dai2019backdoor}, style level (StL)~\cite{qi2021mind, pan2022hidden}, and syntactic level (SyL)~\cite{qi2021hidden, cheng2024syntactic}.
\item[4] / signifies that the validation of such information has not been established, or it has not been performed in the proposed work.
\end{tablenotes}
\end{threeparttable}}
\end{table*}

Table~\ref{tab3} presents a comprehensive summary and comparison of representative backdoor defense methods, as well as a benchmark for uniform evaluation. In this benchmark, defense methods report the ASR ($\Delta$ASR) and CACC ($\Delta$CACC) against backdoor attacks with the `cf' trigger, target label of positive, and a poisoning rate of 10\%. The initial attack CACC and ASR are 92.20\% and 100\%, respectively. To evaluate time complexity, we measure the complexity of detecting a sample during sample filtering, the complexity of purifying a poisoned dataset during sample conversion, the complexity of removing poisoned neurons during model purification, and the complexity of detecting a model during model diagnosis.

\subsubsection{Result Analysis}
In sample inspection, we find that the time complexity of these defenses generally reaches $O(n^2)$, as they need to scan for triggers in each sample~\cite{qi2021onion, shao2021bddr, li2023defending} or perform adversarial perturbations~\cite{gao2019strip, yang2021rap}. Also, these defenses are largely ineffective against both style and syntactic attacks. In terms of defender capabilities, these methods do not require poisoned or validated sets, but they can significantly reduce ASR with the help of internal information, such as logits and attention mechanisms.

In sample conversion, the defender is required to access not only the model and the poisoned dataset, but also the validation dataset~\cite{chen2022expose, he2024seep}. Due to the need for retraining, the time complexity of most approaches reaches $O(n^2)$. Also, representation analysis has proven effective in defending against stealthy triggers. Importantly, these techniques accurately identify backdoor relationships across various dimensions, significantly reducing ASR while maintaining high CACC.

Similarly, the time complexity of defenses in model purification still reaches $O(n^2)$, either due to locating poisoned neurons using a validation dataset~\cite{zhang2023red, liu2018fine, zhang2022fine} or conducting custom retraining strategy on the poisoned dataset~\cite{wu2024acquiring, liu2023shortcuts, graf2024two, tang2023setting}. Compared to fine-pruning, these retraining strategies are effective against four types of attacks. Notably, model purification also achieves promising defense performance compared to dataset purification.

In model diagnosis, the defender typically uses a small validation dataset to generate or convert triggers and then determine whether the model is poisoned. Attention scores and logit outputs from the model also provide valuable information for diagnosis. We observe that these defenses rely on a complex pipeline, where the time complexity is determined by the upper bound or cumulative of all defense steps. For example, clustering typically requires $O(n^2)$~\cite{azizi2021t}, while meta-learning~\cite{xu2021detecting} can reach $O(n^3)$. We argue that $\mathtt{PICCOLO}$~\cite{liu2022piccolo} and constrained optimization~\cite{shen2022constrained} are more practical in terms of both defense type and performance.

\section{Discussion and open challenges}\label{sec5}
In this section, we present a discussion of open issues that deserve further studying, and we offer detailed suggestions for future research directions in the following.

\subsection{Trigger Design}
Although the competitive performance of existing attacks on victim models, few approaches have simultaneously satisfied all objectives of the attacker. Therefore, a viable strategy is to design stealthy triggers (e.g., syntax- or style-based) for APMF and APMP phases to enhance stealthiness. In the AFMT phase, attackers should focus on reducing PPL and increasing USE. Currently, LLMs, as paraphrasing models, are more stealthy for maintaining naturalness and fluency~\cite{xu2023instructions}. Furthermore, attackers should design specific objectives (e.g., dynamic triggers or defense evasion) to inject adaptive backdoors.

\subsection{Extensive Attack Study}
The attacker always poisons the training data with pre-defined triggers and then injects a backdoor into the target model. With triggers in place, the attacker can launch active attacks. However, another insidious method is the passive attack, where benign users unknowingly activate backdoors, since misdirecting a decision model through the actions of many benign users is more powerful than a single attacker. We observe this attack in NLG tasks, such as attacking a pre-defined entity~\cite{yan2023backdooring} and specific instructions~\cite{cheng2024trojanrag}. We argue that it may also be adapted to textual understanding tasks.

Although several studies have compromised NLG models~\cite{xu2021targeted, bagdasaryan2022spinning, wang2021putting, chen2023backdoor}, security threats to other tasks, such as dialogue, creative writing, and freeform question answering, still need to be uncovered. In this survey, we also highlight backdoor vulnerabilities across various scenarios for LLMs. However, we find that existing research primarily exploits word-level triggers to demonstrate the presence of backdoors, rather than fully achieving attackers' objectives. In addition, most studies are evaluated on traditional tasks (e.g., SST-2), which are not sufficiently challenging for LLMs. Given the task-general nature of LLMs, attackers should consider diverse outputs (e.g., contextual consistency) to deceive the alignment mechanism. It is worth noting that efficiency has become a key metric for attacking LLMs. Thus, attackers are equally competitive in reporting time and computational consumption in their follow-up work.

\subsection{Robustness and Effective Defenses}
As discussed in Section~\ref{defense}, existing defenses face numerous limitations. First, most defenses are empirical and demonstrate effectiveness only in specific scenarios. Second, it is a challenge to resist non-insertion attacks. For example, only 7 out of 35 defenses detect style triggers, while only 15 out of 35 defenses detect syntactic triggers. Third, defenders usually focus on incremental improvements for classification tasks while neglecting NLG tasks, particularly in LLMs. Therefore, it is essential to propose an effective and efficient defense for trigger types (e.g., stealthy triggers), task types (e.g., NLG tasks), and objectives (e.g., LLMs). In addition, we suggest that defenders integrate an end-to-end defense framework that not only identifies backdoor models before deployment but also performs sample inspection. Furthermore, benign users should adopt a majority-vote method that randomly selects models from different sources to make collaborative decisions.    


\subsection{Precise Evaluation}
The effectiveness of backdoor attacks depends on how successfully they achieve the attacker's objectives. In general, it is also influenced by trigger types, poisoning rate, and attack strategy. However, many works of backdoor attacks focus on classification, using ASR and CACC as evaluation metrics. In contrast, there are no unified metrics for evaluating NLG tasks, which hinders the development of comprehensive benchmarks in these areas. Moreover, we suggest that attacker report their attack cost, (e.g., time and computation complexity), contributing to a novel track for backdoor attacks. Importantly, existing metrics do not accurately reflect the true impact of a backdoor attack, as ASR can be influenced by external factors such as noisy data, outliers, and semantic shifts.

In contrast, although time complexity is reported in our benchmark, few studies measure it comprehensively or provide detailed evaluation metrics. Also, most defenses use the reduction of attack effectiveness, or anomaly detection metrics as evaluation criteria~\cite{liu2022piccolo}. We argue that the latter is a more appropriate evaluation setup, as defenses can be framed as binary classification tasks on imbalanced datasets. Notably, GPT-4's judgment and auditing capabilities may be promising metrics for LLM defense.

\subsection{Impact Conversion}
Backdoor attacks can be analyzed from both positive and negative perspectives. Hence, these attacks can also bring positive benefits to the NLP community. We highlight several promising research directions as references.
\subsubsection{Watermarking}
Some studies view backdoors as a form of watermarking, used to protect the intellectual property of models and deter unauthorized copying and distribution~\cite{li2023plmmark, gu2022watermarking}. This is because activating the backdoor can be seen as a declaration of model ownership, as the triggers are known only to the provider. 

\subsubsection{Steganography} Many backdoor attack strategies can be applied to enhance the security of information transmission in steganography~\cite{huang2020texthide}. For instance, Yang \textit{et al.}~\cite{yang2023semantic} embed secret data using a semantic-aware information encoding strategy, which is similar to word replacement with synonyms in backdoor attacks. Thus, syntactic structures and linguistic styles can also serve as carriers of secret data.

\subsubsection{Others} Honeypot trapping deliberately uses a backdoor as bait to lure attackers~\cite{le2021sweet}. It is an effective defense against optimization-based triggers (e.g., UOR~\cite{du2023uor}), as adversarial examples are often employed to backdoor samples. Thus, we can utilize the honeypot backdoor to thwart adversarial attacks. Moreover, backdoor implantation provides a practicable option for verifying the deletion of user data~\cite{sommer2020towards}. This is because users poison the data they possess, causing the server to be implanted with a backdoor if it uses such unauthorized data. Indeed, there is no trace of a backdoor if the server performs data deletion. This is particularly relevant for NLP models, whose data originates from diverse sources and is often trained on third-party platforms.

\section{Conclusion}\label{sec6}
Backdoor attacks pose a serious threat to NLP models, while backdoor defenses work to actively mitigate these threats. In this paper, we provide the NLP community with a timely review of backdoor attacks and countermeasures. According to the attackers’ capability and affected stage of the LMs, we outline the attack aims and granularity analysis, and classify the attack surfaces into four categories. Also, we present a comprehensive review of countermeasures against these attacks, structured around the detection objects and their internal goals. Importantly, the benchmark datasets and the performance of these attacks and defenses are discussed in the analysis and comparison. There are many issues in this area that need to be addressed, among which a significant gap between existing attacks and countermeasures still exists.  We hope that this paper provides researchers with a comprehensive overview of backdoor attacks and defenses, and encourages the development of more robust attacks and defenses.

\bibliography{manuscript}

\begin{thebibliography}{100}

\bibitem{li2021hidden}
S.~Li, H.~Liu, T.~Dong, B.~Z.~H. Zhao, M.~Xue, H.~Zhu, and J.~Lu, ``Hidden backdoors in human-centric language models,'' in {\em Proceedings of the 2021 ACM SIGSAC Conference on Computer and Communications Security}, pp.~3123--3140, 2021.

\bibitem{huang2023training}
Y.~Huang, T.~Y. Zhuo, Q.~Xu, H.~Hu, X.~Yuan, and C.~Chen, ``Training-free lexical backdoor attacks on language models,'' in {\em Proceedings of the ACM Web Conference 2023}, pp.~2198--2208, 2023.

\bibitem{sheng2022survey}
X.~Sheng, Z.~Han, P.~Li, and X.~Chang, ``A survey on backdoor attack and defense in natural language processing,'' {\em arXiv preprint arXiv:2211.11958}, 2022.

\bibitem{feng2020securenlp}
Q.~Feng, D.~He, Z.~Liu, H.~Wang, and K.-K.~R. Choo, ``Securenlp: A system for multi-party privacy-preserving natural language processing,'' {\em IEEE Transactions on Information Forensics and Security}, vol.~15, pp.~3709--3721, 2020.

\bibitem{siracusano2019poster}
G.~Siracusano, M.~Trevisan, R.~Gonzalez, and R.~Bifulco, ``Poster: on the application of nlp to discover relationships between malicious network entities,'' in {\em Proceedings of the 2019 ACM SIGSAC Conference on Computer and Communications Security}, pp.~2641--2643, 2019.

\bibitem{gu2017badnets}
T.~Gu, B.~Dolan-Gavitt, and S.~Garg, ``Badnets: Identifying vulnerabilities in the machine learning model supply chain,'' {\em arXiv preprint arXiv:1708.06733}, 2017.

\bibitem{qi2021turn}
F.~Qi, Y.~Yao, S.~Xu, Z.~Liu, and M.~Sun, ``Turn the combination lock: Learnable textual backdoor attacks via word substitution,'' in {\em Proceedings of the 59th Annual Meeting of the Association for Computational Linguistics and the 11th International Joint Conference on Natural Language Processing (Volume 1: Long Papers)}, pp.~4873--4883, 2021.

\bibitem{qi2021hidden}
F.~Qi, M.~Li, Y.~Chen, Z.~Zhang, Z.~Liu, Y.~Wang, and M.~Sun, ``Hidden killer: Invisible textual backdoor attacks with syntactic trigger,'' in {\em Proceedings of the 59th Annual Meeting of the Association for Computational Linguistics and the 11th International Joint Conference on Natural Language Processing (Volume 1: Long Papers)}, pp.~443--453, 2021.

\bibitem{qi2021mind}
F.~Qi, Y.~Chen, X.~Zhang, M.~Li, Z.~Liu, and M.~Sun, ``Mind the style of text! adversarial and backdoor attacks based on text style transfer,'' in {\em Proceedings of the 2021 Conference on Empirical Methods in Natural Language Processing}, pp.~4569--4580, 2021.

\bibitem{dai2019backdoor}
J.~Dai, C.~Chen, and Y.~Li, ``A backdoor attack against lstm-based text classification systems,'' {\em IEEE Access}, vol.~7, pp.~138872--138878, 2019.

\bibitem{chen2021badnl}
X.~Chen, A.~Salem, D.~Chen, M.~Backes, S.~Ma, Q.~Shen, Z.~Wu, and Y.~Zhang, ``Badnl: Backdoor attacks against nlp models with semantic-preserving improvements,'' in {\em 37th Annual Computer Security Applications Conference, ACSAC 2021}, pp.~554--569, Association for Computing Machinery, 2021.

\bibitem{shen2022rethink}
L.~Shen, H.~Jiang, L.~Liu, and S.~Shi, ``Rethink the evaluation for attack strength of backdoor attacks in natural language processing,'' {\em arXiv preprint arXiv:2201.02993}, 2022.

\bibitem{zhang2023red}
Z.~Zhang, G.~Xiao, Y.~Li, T.~Lv, F.~Qi, Z.~Liu, Y.~Wang, X.~Jiang, and M.~Sun, ``Red alarm for pre-trained models: Universal vulnerability to neuron-level backdoor attacks,'' {\em Machine Intelligence Research}, pp.~1--14, 2023.

\bibitem{shen2021backdoor}
L.~Shen, S.~Ji, X.~Zhang, J.~Li, J.~Chen, J.~Shi, C.~Fang, J.~Yin, and T.~Wang, ``Backdoor pre-trained models can transfer to all,'' in {\em Proceedings of the 2021 ACM SIGSAC Conference on Computer and Communications Security}, pp.~3141--3158, 2021.

\bibitem{zulqarnain2020comparative}
M.~Zulqarnain, R.~Ghazali, Y.~M.~M. Hassim, and M.~Rehan, ``A comparative review on deep learning models for text classification,'' {\em Indones. J. Electr. Eng. Comput. Sci}, vol.~19, no.~1, pp.~325--335, 2020.

\bibitem{tan2020neural}
Z.~Tan, S.~Wang, Z.~Yang, G.~Chen, X.~Huang, M.~Sun, and Y.~Liu, ``Neural machine translation: A review of methods, resources, and tools,'' {\em AI Open}, vol.~1, pp.~5--21, 2020.

\bibitem{alwaneen2022arabic}
T.~H. Alwaneen, A.~M. Azmi, H.~A. Aboalsamh, E.~Cambria, and A.~Hussain, ``Arabic question answering system: a survey,'' {\em Artificial Intelligence Review}, pp.~1--47, 2022.

\bibitem{qi2021onion}
F.~Qi, Y.~Chen, M.~Li, Y.~Yao, Z.~Liu, and M.~Sun, ``Onion: A simple and effective defense against textual backdoor attacks,'' in {\em Proceedings of the 2021 Conference on Empirical Methods in Natural Language Processing}, pp.~9558--9566, 2021.

\bibitem{gao2019strip}
Y.~Gao, C.~Xu, D.~Wang, S.~Chen, D.~C. Ranasinghe, and S.~Nepal, ``Strip: A defence against trojan attacks on deep neural networks,'' in {\em Proceedings of the 35th Annual Computer Security Applications Conference}, pp.~113--125, 2019.

\bibitem{azizi2021t}
A.~Azizi, I.~A. Tahmid, A.~Waheed, N.~Mangaokar, J.~Pu, M.~Javed, C.~K. Reddy, and B.~Viswanath, ``T-miner: A generative approach to defend against trojan attacks on dnn-based text classification,'' {\em arXiv preprint arXiv:2103.04264}, 2021.

\bibitem{li2022backdoors}
S.~Li, T.~Dong, B.~Z.~H. Zhao, M.~Xue, S.~Du, and H.~Zhu, ``Backdoors against natural language processing: A review,'' {\em IEEE Security \& Privacy}, vol.~20, no.~5, pp.~50--59, 2022.

\bibitem{zhao2024survey}
S.~Zhao, M.~Jia, Z.~Guo, L.~Gan, J.~Fu, Y.~Feng, F.~Pan, and L.~A. Tuan, ``A survey of backdoor attacks and defenses on large language models: Implications for security measures,'' {\em arXiv preprint arXiv:2406.06852}, 2024.

\bibitem{devlin2018bert}
J.~Devlin, M.-W. Chang, K.~Lee, and K.~Toutanova, ``Bert: Pre-training of deep bidirectional transformers for language understanding,'' {\em arXiv preprint arXiv:1810.04805}, 2018.

\bibitem{zhao2024universal}
S.~Zhao, M.~Jia, L.~A. Tuan, F.~Pan, and J.~Wen, ``Universal vulnerabilities in large language models: Backdoor attacks for in-context learning,'' {\em arXiv preprint arXiv:2401.05949}, 2024.

\bibitem{xiang2023badchain}
Z.~Xiang, F.~Jiang, Z.~Xiong, B.~Ramasubramanian, R.~Poovendran, and B.~Li, ``Badchain: Backdoor chain-of-thought prompting for large language models,'' in {\em NeurIPS 2023 Workshop on Backdoors in Deep Learning-The Good, the Bad, and the Ugly}, 2023.

\bibitem{chenbadpre}
K.~Chen, Y.~Meng, X.~Sun, S.~Guo, T.~Zhang, J.~Li, and C.~Fan, ``Badpre: Task-agnostic backdoor attacks to pre-trained nlp foundation models,'' in {\em International Conference on Learning Representations}, 2021.

\bibitem{xu2021targeted}
C.~Xu, J.~Wang, Y.~Tang, F.~Guzm{\'a}n, B.~I. Rubinstein, and T.~Cohn, ``A targeted attack on black-box neural machine translation with parallel data poisoning,'' in {\em Proceedings of the web conference 2021}, pp.~3638--3650, 2021.

\bibitem{xue2024trojllm}
J.~Xue, M.~Zheng, T.~Hua, Y.~Shen, Y.~Liu, L.~B{\"o}l{\"o}ni, and Q.~Lou, ``Trojllm: A black-box trojan prompt attack on large language models,'' {\em Advances in Neural Information Processing Systems}, vol.~36, 2024.

\bibitem{cuiunified}
G.~Cui, L.~Yuan, B.~He, Y.~Chen, Z.~Liu, and M.~Sun, ``A unified evaluation of textual backdoor learning: Frameworks and benchmarks,'' in {\em Thirty-sixth Conference on Neural Information Processing Systems Datasets and Benchmarks Track}, 2022.

\bibitem{kurita2020weight}
K.~Kurita, P.~Michel, and G.~Neubig, ``Weight poisoning attacks on pretrained models,'' in {\em Proceedings of the 58th Annual Meeting of the Association for Computational Linguistics}, pp.~2793--2806, 2020.

\bibitem{li2021backdoor}
L.~Li, D.~Song, X.~Li, J.~Zeng, R.~Ma, and X.~Qiu, ``Backdoor attacks on pre-trained models by layerwise weight poisoning,'' in {\em Proceedings of the 2021 Conference on Empirical Methods in Natural Language Processing}, pp.~3023--3032, 2021.

\bibitem{gao2020backdoor}
Y.~Gao, B.~G. Doan, Z.~Zhang, S.~Ma, J.~Zhang, A.~Fu, S.~Nepal, and H.~Kim, ``Backdoor attacks and countermeasures on deep learning: A comprehensive review,'' {\em arXiv preprint arXiv:2007.10760}, 2020.

\bibitem{yang2021careful}
W.~Yang, L.~Li, Z.~Zhang, X.~Ren, X.~Sun, and B.~He, ``Be careful about poisoned word embeddings: Exploring the vulnerability of the embedding layers in nlp models,'' in {\em Proceedings of the 2021 Conference of the North American Chapter of the Association for Computational Linguistics: Human Language Technologies}, pp.~2048--2058, 2021.

\bibitem{zhang2021neural}
Z.~Zhang, X.~Ren, Q.~Su, X.~Sun, and B.~He, ``Neural network surgery: Injecting data patterns into pre-trained models with minimal instance-wise side effects,'' in {\em Proceedings of the 2021 Conference of the North American Chapter of the Association for Computational Linguistics: Human Language Technologies}, pp.~5453--5466, 2021.

\bibitem{du2023uor}
W.~Du, P.~Li, B.~Li, H.~Zhao, and G.~Liu, ``Uor: Universal backdoor attacks on pre-trained language models,'' {\em arXiv preprint arXiv:2305.09574}, 2023.

\bibitem{cheng2024syntactic}
P.~Cheng, W.~Du, Z.~Wu, F.~Zhang, L.~Chen, and G.~Liu, ``Syntactic ghost: An imperceptible general-purpose backdoor attacks on pre-trained language models,'' {\em arXiv preprint arXiv:2402.18945}, 2024.

\bibitem{xu2022exploring}
L.~Xu, Y.~Chen, G.~Cui, H.~Gao, and Z.~Liu, ``Exploring the universal vulnerability of prompt-based learning paradigm,'' in {\em Findings of the Association for Computational Linguistics: NAACL 2022}, pp.~1799--1810, 2022.

\bibitem{zhao2023prompt}
S.~Zhao, J.~Wen, L.~A. Tuan, J.~Zhao, and J.~Fu, ``Prompt as triggers for backdoor attack: Examining the vulnerability in language models,'' {\em arXiv preprint arXiv:2305.01219}, 2023.

\bibitem{du2022ppt}
W.~Du, Y.~Zhao, B.~Li, G.~Liu, and S.~Wang, ``Ppt: Backdoor attacks on pre-trained models via poisoned prompt tuning,'' in {\em Proceedings of the Thirty-First International Joint Conference on Artificial Intelligence, IJCAI-22}, pp.~680--686, 2022.

\bibitem{cai2022badprompt}
X.~Cai, H.~Xu, S.~Xu, Y.~Zhang, {\em et~al.}, ``Badprompt: Backdoor attacks on continuous prompts,'' {\em Advances in Neural Information Processing Systems}, vol.~35, pp.~37068--37080, 2022.

\bibitem{gu2023gradient}
N.~Gu, P.~Fu, X.~Liu, Z.~Liu, Z.~Lin, and W.~Wang, ``A gradient control method for backdoor attacks on parameter-efficient tuning,'' in {\em Proceedings of the 61st Annual Meeting of the Association for Computational Linguistics (Volume 1: Long Papers)}, pp.~3508--3520, 2023.

\bibitem{kwon2021textual}
H.~Kwon and S.~Lee, ``Textual backdoor attack for the text classification system,'' {\em Security and Communication Networks}, vol.~2021, pp.~1--11, 2021.

\bibitem{lu2022attack}
H.-y. Lu, C.~Fan, J.~Yang, C.~Hu, W.~Fang, and X.-j. Wu, ``Where to attack: A dynamic locator model for backdoor attack in text classifications,'' in {\em Proceedings of the 29th International Conference on Computational Linguistics}, pp.~984--993, 2022.

\bibitem{chen-etal-2022-textual}
Y.~Chen, F.~Qi, H.~Gao, Z.~Liu, and M.~Sun, ``Textual backdoor attacks can be more harmful via two simple tricks,'' in {\em Proceedings of the 2022 Conference on Empirical Methods in Natural Language Processing}, (Abu Dhabi, United Arab Emirates), pp.~11215--11221, Association for Computational Linguistics, Dec. 2022.

\bibitem{bagdasaryan2022spinning}
E.~Bagdasaryan and V.~Shmatikov, ``Spinning language models: Risks of propaganda-as-a-service and countermeasures,'' in {\em 2022 IEEE Symposium on Security and Privacy (SP)}, pp.~769--786, IEEE, 2022.

\bibitem{yang2021rethinking}
W.~Yang, Y.~Lin, P.~Li, J.~Zhou, and X.~Sun, ``Rethinking stealthiness of backdoor attack against nlp models,'' in {\em Proceedings of the 59th Annual Meeting of the Association for Computational Linguistics and the 11th International Joint Conference on Natural Language Processing (Volume 1: Long Papers)}, pp.~5543--5557, 2021.

\bibitem{gan2022triggerless}
L.~Gan, J.~Li, T.~Zhang, X.~Li, Y.~Meng, F.~Wu, Y.~Yang, S.~Guo, and C.~Fan, ``Triggerless backdoor attack for nlp tasks with clean labels,'' in {\em Proceedings of the 2022 Conference of the North American Chapter of the Association for Computational Linguistics: Human Language Technologies}, pp.~2942--2952, 2022.

\bibitem{liu2023trojtext}
Y.~Liu, B.~Feng, and Q.~Lou, ``Trojtext: Test-time invisible textual trojan insertion,'' {\em arXiv preprint arXiv:2303.02242}, 2023.

\bibitem{salem2021badnl}
X.~C.~A. Salem and M.~Zhang, ``Badnl: Backdoor attacks against nlp models,'' in {\em ICML 2021 Workshop on Adversarial Machine Learning}, 2021.

\bibitem{pan2022hidden}
X.~Pan, M.~Zhang, B.~Sheng, J.~Zhu, and M.~Yang, ``Hidden trigger backdoor attack on $\{$NLP$\}$ models via linguistic style manipulation,'' in {\em 31st USENIX Security Symposium (USENIX Security 22)}, pp.~3611--3628, 2022.

\bibitem{li2023chatgpt}
J.~Li, Y.~Yang, Z.~Wu, V.~Vydiswaran, and C.~Xiao, ``Chatgpt as an attack tool: Stealthy textual backdoor attack via blackbox generative model trigger,'' {\em arXiv preprint arXiv:2304.14475}, 2023.

\bibitem{shao2022triggers}
K.~Shao, Y.~Zhang, J.~Yang, X.~Li, and H.~Liu, ``The triggers that open the nlp model backdoors are hidden in the adversarial samples,'' {\em Computers \& Security}, vol.~118, p.~102730, 2022.

\bibitem{garg2020can}
S.~Garg, A.~Kumar, V.~Goel, and Y.~Liang, ``Can adversarial weight perturbations inject neural backdoors,'' in {\em Proceedings of the 29th ACM International Conference on Information \& Knowledge Management}, pp.~2029--2032, 2020.

\bibitem{maqsood2022backdoor}
S.~M. Maqsood, V.~M. Ceron, and A.~GowthamKrishna, ``Backdoor attack against nlp models with robustness-aware perturbation defense,'' {\em arXiv preprint arXiv:2204.05758}, 2022.

\bibitem{zhou2023backdoor}
X.~Zhou, J.~Li, T.~Zhang, L.~Lyu, M.~Yang, and J.~He, ``Backdoor attacks with input-unique triggers in nlp,'' {\em arXiv preprint arXiv:2303.14325}, 2023.

\bibitem{gupta2023adversarial}
A.~Gupta and A.~Krishna, ``Adversarial clean label backdoor attacks and defenses on text classification systems,'' {\em arXiv preprint arXiv:2305.19607}, 2023.

\bibitem{chen2022kallima}
X.~Chen, Y.~Dong, Z.~Sun, S.~Zhai, Q.~Shen, and Z.~Wu, ``Kallima: A clean-label framework for textual backdoor attacks,'' in {\em Computer Security--ESORICS 2022: 27th European Symposium on Research in Computer Security, Copenhagen, Denmark, September 26--30, 2022, Proceedings, Part I}, pp.~447--466, Springer, 2022.

\bibitem{yan2023bite}
J.~Yan, V.~Gupta, and X.~Ren, ``Bite: Textual backdoor attacks with iterative trigger injection,'' 2023.

\bibitem{du2024backdoor}
W.~Du, T.~Ju, G.~Ren, G.~Li, and G.~Liu, ``Backdoor nlp models via ai-generated text,'' in {\em Proceedings of the 2024 Joint International Conference on Computational Linguistics, Language Resources and Evaluation (LREC-COLING 2024)}, pp.~2067--2079, 2024.

\bibitem{du2024nws}
W.~Du, T.~Yuan, H.~Zhao, and G.~Liu, ``Nws: Natural textual backdoor attacks via word substitution,'' in {\em ICASSP 2024-2024 IEEE International Conference on Acoustics, Speech and Signal Processing (ICASSP)}, pp.~4680--4684, IEEE, 2024.

\bibitem{xu2023instructions}
J.~Xu, M.~D. Ma, F.~Wang, C.~Xiao, and M.~Chen, ``Instructions as backdoors: Backdoor vulnerabilities of instruction tuning for large language models,'' {\em arXiv preprint arXiv:2305.14710}, 2023.

\bibitem{wei2023bdmmt}
J.~Wei, M.~Fan, W.~Jiao, W.~Jin, and T.~Liu, ``Bdmmt: Backdoor sample detection for language models through model mutation testing,'' {\em arXiv preprint arXiv:2301.10412}, 2023.

\bibitem{qiang2024learning}
Y.~Qiang, X.~Zhou, S.~Z. Zade, M.~A. Roshani, D.~Zytko, and D.~Zhu, ``Learning to poison large language models during instruction tuning,'' {\em arXiv preprint arXiv:2402.13459}, 2024.

\bibitem{yao2023poisonprompt}
H.~Yao, J.~Lou, and Z.~Qin, ``Poisonprompt: Backdoor attack on prompt-based large language models,'' {\em arXiv preprint arXiv:2310.12439}, 2023.

\bibitem{zhang2024rapid}
R.~Zhang, H.~Li, R.~Wen, W.~Jiang, Y.~Zhang, M.~Backes, Y.~Shen, and Y.~Zhang, ``Rapid adoption, hidden risks: The dual impact of large language model customization,'' {\em arXiv preprint arXiv:2402.09179}, 2024.

\bibitem{sheng2023punctuation}
X.~Sheng, Z.~Li, Z.~Han, X.~Chang, and P.~Li, ``Punctuation matters! stealthy backdoor attack for language models,'' in {\em CCF International Conference on Natural Language Processing and Chinese Computing}, pp.~524--536, Springer, 2023.

\bibitem{li2024leverage}
X.~Li, X.~Lu, and P.~Li, ``Leverage nlp models against other nlp models: Two invisible feature space backdoor attacks,'' {\em IEEE Transactions on Reliability}, 2024.

\bibitem{zhao2024exploring}
S.~Zhao, L.~A. Tuan, J.~Fu, J.~Wen, and W.~Luo, ``Exploring clean label backdoor attacks and defense in language models,'' {\em IEEE/ACM Transactions on Audio, Speech, and Language Processing}, 2024.

\bibitem{libadedit}
Y.~Li, T.~Li, K.~Chen, J.~Zhang, S.~Liu, W.~Wang, T.~Zhang, and Y.~Liu, ``Badedit: Backdooring large language models by model editing,'' in {\em The Twelfth International Conference on Learning Representations}, 2024.

\bibitem{qiu2024megen}
J.~Qiu, X.~Ma, Z.~Zhang, and H.~Zhao, ``Megen: Generative backdoor in large language models via model editing,'' {\em arXiv preprint arXiv:2408.10722}, 2024.

\bibitem{cheng2024transferring}
P.~Cheng, Z.~Wu, T.~Ju, W.~Du, and Z.~Z.~G. Liu, ``Transferring backdoors between large language models by knowledge distillation,'' {\em arXiv preprint arXiv:2408.09878}, 2024.

\bibitem{tan2023target}
Z.~Tan, Q.~Chen, Y.~Huang, and C.~Liang, ``Target: Template-transferable backdoor attack against prompt-based nlp models via gpt4,'' {\em arXiv preprint arXiv:2311.17429}, 2023.

\bibitem{nie2024trojfm}
Y.~Nie, Y.~Wang, J.~Jia, M.~J. De~Lucia, N.~D. Bastian, W.~Guo, and D.~Song, ``Trojfm: Resource-efficient backdoor attacks against very large foundation models,'' {\em arXiv preprint arXiv:2405.16783}, 2024.

\bibitem{shao2021bddr}
K.~Shao, J.~Yang, Y.~Ai, H.~Liu, and Y.~Zhang, ``Bddr: An effective defense against textual backdoor attacks,'' {\em Computers \& Security}, vol.~110, p.~102433, 2021.

\bibitem{he2023imbert}
X.~He, J.~Wang, B.~Rubinstein, and T.~Cohn, ``Imbert: Making bert immune to insertion-based backdoor attacks,'' {\em arXiv preprint arXiv:2305.16503}, 2023.

\bibitem{yan2024parafuzz}
L.~Yan, Z.~Zhang, G.~Tao, K.~Zhang, X.~Chen, G.~Shen, and X.~Zhang, ``Parafuzz: An interpretability-driven technique for detecting poisoned samples in nlp,'' {\em Advances in Neural Information Processing Systems}, vol.~36, 2024.

\bibitem{zhao2024defending}
S.~Zhao, L.~Gan, L.~A. Tuan, J.~Fu, L.~Lyu, M.~Jia, and J.~Wen, ``Defending against weight-poisoning backdoor attacks for parameter-efficient fine-tuning,'' {\em arXiv preprint arXiv:2402.12168}, 2024.

\bibitem{yang2021rap}
W.~Yang, Y.~Lin, P.~Li, J.~Zhou, and X.~Sun, ``Rap: Robustness-aware perturbations for defending against backdoor attacks on nlp models,'' in {\em Proceedings of the 2021 Conference on Empirical Methods in Natural Language Processing}, pp.~8365--8381, 2021.

\bibitem{le2021sweet}
T.~L. Le, N.~P. Park, and D.~Lee, ``A sweet rabbit hole by darcy: Using honeypots to detect universal trigger's adversarial attacks,'' in {\em 59th Annual Meeting of the Association for Comp. Linguistics (ACL)}, 2021.

\bibitem{xian2023unified}
X.~Xian, G.~Wang, J.~Srinivasa, A.~Kundu, X.~Bi, M.~Hong, and J.~Ding, ``A unified detection framework for inference-stage backdoor defenses,'' {\em Advances in Neural Information Processing Systems}, vol.~36, pp.~7867--7894, 2023.

\bibitem{yi2024badacts}
B.~Yi, S.~Chen, Y.~Li, T.~Li, B.~Zhang, and Z.~Liu, ``Badacts: A universal backdoor defense in the activation space,'' {\em arXiv preprint arXiv:2405.11227}, 2024.

\bibitem{li2021bfclass}
Z.~Li, D.~Mekala, C.~Dong, and J.~Shang, ``Bfclass: A backdoor-free text classification framework,'' in {\em Findings of the Association for Computational Linguistics: EMNLP 2021}, pp.~444--453, 2021.

\bibitem{fan2021text}
M.~Fan, Z.~Si, X.~Xie, Y.~Liu, and T.~Liu, ``Text backdoor detection using an interpretable rnn abstract model,'' {\em IEEE Transactions on Information Forensics and Security}, vol.~16, pp.~4117--4132, 2021.

\bibitem{chen2021mitigating}
C.~Chen and J.~Dai, ``Mitigating backdoor attacks in lstm-based text classification systems by backdoor keyword identification,'' {\em Neurocomputing}, vol.~452, pp.~253--262, 2021.

\bibitem{li2023defending}
J.~Li, Z.~Wu, W.~Ping, C.~Xiao, and V.~Vydiswaran, ``Defending against insertion-based textual backdoor attacks via attribution,'' {\em arXiv preprint arXiv:2305.02394}, 2023.

\bibitem{sun2021general}
X.~Sun, J.~Li, X.~Li, Z.~Wang, T.~Zhang, H.~Qiu, F.~Wu, and C.~Fan, ``A general framework for defending against backdoor attacks via influence graph,'' {\em arXiv preprint arXiv:2111.14309}, 2021.

\bibitem{jin2022wedef}
L.~Jin, Z.~Wang, and J.~Shang, ``Wedef: Weakly supervised backdoor defense for text classification,'' in {\em Proceedings of the 2022 Conference on Empirical Methods in Natural Language Processing}, pp.~11614--11626, 2022.

\bibitem{he2023mitigating}
X.~He, Q.~Xu, J.~Wang, B.~Rubinstein, and T.~Cohn, ``Mitigating backdoor poisoning attacks through the lens of spurious correlation,'' {\em arXiv preprint arXiv:2305.11596}, 2023.

\bibitem{chen2022expose}
S.~Chen, W.~Yang, Z.~Zhang, X.~Bi, and X.~Sun, ``Expose backdoors on the way: A feature-based efficient defense against textual backdoor attacks,'' in {\em Findings of the Association for Computational Linguistics: EMNLP 2022}, pp.~668--683, 2022.

\bibitem{xi2024defending}
Z.~Xi, T.~Du, C.~Li, R.~Pang, S.~Ji, J.~Chen, F.~Ma, and T.~Wang, ``Defending pre-trained language models as few-shot learners against backdoor attacks,'' {\em Advances in Neural Information Processing Systems}, vol.~36, 2024.

\bibitem{he2024seep}
X.~He, Q.~Xu, J.~Wang, B.~I. Rubinstein, and T.~Cohn, ``Seep: Training dynamics grounds latent representation search for mitigating backdoor poisoning attacks,'' {\em arXiv preprint arXiv:2405.11575}, 2024.

\bibitem{zhang2022fine}
Z.~Zhang, L.~Lyu, X.~Ma, C.~Wang, and X.~Sun, ``Fine-mixing: Mitigating backdoors in fine-tuned language models,'' in {\em Findings of the Association for Computational Linguistics: EMNLP 2022}, pp.~355--372, 2022.

\bibitem{zhang2023diffusion}
Z.~Zhang, D.~Chen, H.~Zhou, F.~Meng, J.~Zhou, and X.~Sun, ``Diffusion theory as a scalpel: Detecting and purifying poisonous dimensions in pre-trained language models caused by backdoor or bias,'' {\em arXiv preprint arXiv:2305.04547}, 2023.

\bibitem{zhu2022moderate}
B.~Zhu, Y.~Qin, G.~Cui, Y.~Chen, W.~Zhao, C.~Fu, Y.~Deng, Z.~Liu, J.~Wang, W.~Wu, {\em et~al.}, ``Moderate-fitting as a natural backdoor defender for pre-trained language models,'' {\em Advances in Neural Information Processing Systems}, vol.~35, pp.~1086--1099, 2022.

\bibitem{liu2023maximum}
Z.~Liu, B.~Shen, Z.~Lin, F.~Wang, and W.~Wang, ``Maximum entropy loss, the silver bullet targeting backdoor attacks in pre-trained language models,'' in {\em Findings of the Association for Computational Linguistics: ACL 2023}, pp.~3850--3868, 2023.

\bibitem{wu2024acquiring}
Z.~Wu, Z.~Zhang, P.~Cheng, and G.~Liu, ``Acquiring clean language models from backdoor poisoned datasets by downscaling frequency space,'' {\em arXiv preprint arXiv:2402.12026}, 2024.

\bibitem{liu2023shortcuts}
Q.~Liu, F.~Wang, C.~Xiao, and M.~Chen, ``From shortcuts to triggers: Backdoor defense with denoised poe,'' {\em arXiv preprint arXiv:2305.14910}, 2023.

\bibitem{graf2024two}
V.~Graf, Q.~Liu, and M.~Chen, ``Two heads are better than one: Nested poe for robust defense against multi-backdoors,'' {\em arXiv preprint arXiv:2404.02356}, 2024.

\bibitem{tang2023setting}
R.~R. Tang, J.~Yuan, Y.~Li, Z.~Liu, R.~Chen, and X.~Hu, ``Setting the trap: Capturing and defeating backdoors in pretrained language models through honeypots,'' {\em Advances in Neural Information Processing Systems}, vol.~36, pp.~73191--73210, 2023.

\bibitem{shen2022constrained}
G.~Shen, Y.~Liu, G.~Tao, Q.~Xu, Z.~Zhang, S.~An, S.~Ma, and X.~Zhang, ``Constrained optimization with dynamic bound-scaling for effective nlp backdoor defense,'' in {\em International Conference on Machine Learning}, pp.~19879--19892, PMLR, 2022.

\bibitem{lyu2022study}
W.~Lyu, S.~Zheng, T.~Ma, and C.~Chen, ``A study of the attention abnormality in trojaned berts,'' in {\em Annual Conference of the North American Chapter of the Association for Computational Linguistics}, 2022.

\bibitem{lyu2024task}
W.~Lyu, X.~Lin, S.~Zheng, L.~Pang, H.~Ling, S.~Jha, and C.~Chen, ``Task-agnostic detector for insertion-based backdoor attacks,'' {\em arXiv preprint arXiv:2403.17155}, 2024.

\bibitem{zeng2024clibe}
R.~Zeng, X.~Chen, Y.~Pu, X.~Zhang, T.~Du, and S.~Ji, ``Clibe: Detecting dynamic backdoors in transformer-based nlp models,'' {\em arXiv preprint arXiv:2409.01193}, 2024.

\bibitem{zhang2021trojaning}
X.~Zhang, Z.~Zhang, S.~Ji, and T.~Wang, ``Trojaning language models for fun and profit,'' in {\em 2021 IEEE European Symposium on Security and Privacy (EuroS\&P)}, pp.~179--197, IEEE, 2021.

\bibitem{xu2021detecting}
X.~Xu, Q.~Wang, H.~Li, N.~Borisov, C.~A. Gunter, and B.~Li, ``Detecting ai trojans using meta neural analysis,'' in {\em 2021 IEEE Symposium on Security and Privacy (SP)}, pp.~103--120, IEEE, 2021.

\bibitem{wallace2021concealed}
E.~Wallace, T.~Zhao, S.~Feng, and S.~Singh, ``Concealed data poisoning attacks on nlp models,'' in {\em Proceedings of the 2021 Conference of the North American Chapter of the Association for Computational Linguistics: Human Language Technologies}, pp.~139--150, 2021.

\bibitem{wang2021putting}
J.~Wang, C.~Xu, F.~Guzm{\'a}n, A.~El-Kishky, Y.~Tang, B.~Rubinstein, and T.~Cohn, ``Putting words into the system’s mouth: A targeted attack on neural machine translation using monolingual data poisoning,'' in {\em Findings of the Association for Computational Linguistics: ACL-IJCNLP 2021}, pp.~1463--1473, 2021.

\bibitem{chen2023backdoor}
L.~Chen, M.~Cheng, and H.~Huang, ``Backdoor learning on sequence to sequence models,'' {\em arXiv preprint arXiv:2305.02424}, 2023.

\bibitem{sun2023defending}
X.~Sun, X.~Li, Y.~Meng, X.~Ao, L.~Lyu, J.~Li, and T.~Zhang, ``Defending against backdoor attacks in natural language generation,'' in {\em Proceedings of the AAAI Conference on Artificial Intelligence}, vol.~37, pp.~5257--5265, 2023.

\bibitem{yang2024stealthy}
Z.~Yang, B.~Xu, J.~M. Zhang, H.~J. Kang, J.~Shi, J.~He, and D.~Lo, ``Stealthy backdoor attack for code models,'' {\em IEEE Transactions on Software Engineering}, 2024.

\bibitem{jiang2023forcing}
S.~Jiang, S.~Kadhe, Y.~Zhou, L.~Cai, and N.~Baracaldo, ``Forcing generative models to degenerate ones: The power of data poisoning attacks,'' in {\em NeurIPS 2023 Workshop on Backdoors in Deep Learning-The Good, the Bad, and the Ugly}, 2023.

\bibitem{dong2023unleashing}
T.~Dong, G.~Chen, S.~Li, M.~Xue, R.~Holland, Y.~Meng, Z.~Liu, and H.~Zhu, ``Unleashing cheapfakes through trojan plugins of large language models,'' {\em arXiv preprint arXiv:2312.00374}, 2023.

\bibitem{long2024backdoor}
Q.~Long, Y.~Deng, L.~Gan, W.~Wang, and S.~J. Pan, ``Backdoor attacks on dense passage retrievers for disseminating misinformation,'' {\em arXiv preprint arXiv:2402.13532}, 2024.

\bibitem{cao2023stealthy}
Y.~Cao, B.~Cao, and J.~Chen, ``Stealthy and persistent unalignment on large language models via backdoor injections,'' {\em arXiv preprint arXiv:2312.00027}, 2023.

\bibitem{wang2023backdoor}
H.~Wang and K.~Shu, ``Backdoor activation attack: Attack large language models using activation steering for safety-alignment,'' {\em arXiv preprint arXiv:2311.09433}, 2023.

\bibitem{li2024chain}
X.~Li, Y.~Zhang, R.~Lou, C.~Wu, and J.~Wang, ``Chain-of-scrutiny: Detecting backdoor attacks for large language models,'' {\em arXiv preprint arXiv:2406.05948}, 2024.

\bibitem{li2024backdoor}
H.~Li, Y.~Chen, Z.~Zheng, Q.~Hu, C.~Chan, H.~Liu, and Y.~Song, ``Backdoor removal for generative large language models,'' {\em arXiv preprint arXiv:2405.07667}, 2024.

\bibitem{yang2024watch}
W.~Yang, X.~Bi, Y.~Lin, S.~Chen, J.~Zhou, and X.~Sun, ``Watch out for your agents! investigating backdoor threats to llm-based agents,'' {\em arXiv preprint arXiv:2402.11208}, 2024.

\bibitem{yan2023backdooring}
J.~Yan, V.~Yadav, S.~Li, L.~Chen, Z.~Tang, H.~Wang, V.~Srinivasan, X.~Ren, and H.~Jin, ``Backdooring instruction-tuned large language models with virtual prompt injection,'' in {\em NeurIPS 2023 Workshop on Backdoors in Deep Learning-The Good, the Bad, and the Ugly}, 2023.

\bibitem{wang2024badagent}
Y.~Wang, D.~Xue, S.~Zhang, and S.~Qian, ``Badagent: Inserting and activating backdoor attacks in llm agents,'' {\em arXiv preprint arXiv:2406.03007}, 2024.

\bibitem{huang2023composite}
H.~Huang, Z.~Zhao, M.~Backes, Y.~Shen, and Y.~Zhang, ``Composite backdoor attacks against large language models,'' {\em arXiv preprint arXiv:2310.07676}, 2023.

\bibitem{liu2024compromising}
A.~Liu, Y.~Zhou, X.~Liu, T.~Zhang, S.~Liang, J.~Wang, Y.~Pu, T.~Li, J.~Zhang, W.~Zhou, {\em et~al.}, ``Compromising embodied agents with contextual backdoor attacks,'' {\em arXiv preprint arXiv:2408.02882}, 2024.

\bibitem{li2023imtm}
Z.~Li, P.~Li, X.~Sheng, C.~Yin, and L.~Zhou, ``Imtm: Invisible multi-trigger multimodal backdoor attack,'' in {\em CCF International Conference on Natural Language Processing and Chinese Computing}, pp.~533--545, Springer, 2023.

\bibitem{lu2024test}
D.~Lu, T.~Pang, C.~Du, Q.~Liu, X.~Yang, and M.~Lin, ``Test-time backdoor attacks on multimodal large language models,'' {\em arXiv preprint arXiv:2402.08577}, 2024.

\bibitem{jiao2024exploring}
R.~Jiao, S.~Xie, J.~Yue, T.~Sato, L.~Wang, Y.~Wang, Q.~A. Chen, and Q.~Zhu, ``Exploring backdoor attacks against large language model-based decision making,'' {\em arXiv preprint arXiv:2405.20774}, 2024.

\bibitem{zhu2024seer}
L.~Zhu, R.~Ning, J.~Li, C.~Xin, and H.~Wu, ``Seer: Backdoor detection for vision-language models through searching target text and image trigger jointly,'' in {\em Proceedings of the AAAI Conference on Artificial Intelligence}, vol.~38, pp.~7766--7774, 2024.

\bibitem{sur2023tijo}
I.~Sur, K.~Sikka, M.~Walmer, K.~Koneripalli, A.~Roy, X.~Lin, A.~Divakaran, and S.~Jha, ``Tijo: Trigger inversion with joint optimization for defending multimodal backdoored models,'' in {\em Proceedings of the IEEE/CVF International Conference on Computer Vision}, pp.~165--175, 2023.

\bibitem{papineni2002bleu}
K.~Papineni, S.~Roukos, T.~Ward, and W.-J. Zhu, ``Bleu: a method for automatic evaluation of machine translation,'' in {\em Proceedings of the 40th annual meeting of the Association for Computational Linguistics}, pp.~311--318, 2002.

\bibitem{lin2004rouge}
C.-Y. Lin, ``Rouge: A package for automatic evaluation of summaries,'' in {\em Text summarization branches out}, pp.~74--81, 2004.

\bibitem{reimers2019sentence}
N.~Reimers and I.~Gurevych, ``Sentence-bert: Sentence embeddings using siamese bert-networks,'' in {\em Proceedings of the 2019 Conference on Empirical Methods in Natural Language Processing and the 9th International Joint Conference on Natural Language Processing (EMNLP-IJCNLP)}, pp.~3982--3992, 2019.

\bibitem{cer2018universal}
D.~Cer, Y.~Yang, S.-y. Kong, N.~Hua, N.~Limtiaco, R.~S. John, N.~Constant, M.~Guajardo-Cespedes, S.~Yuan, C.~Tar, {\em et~al.}, ``Universal sentence encoder for english,'' in {\em Proceedings of the 2018 conference on empirical methods in natural language processing: system demonstrations}, pp.~169--174, 2018.

\bibitem{peft}
S.~Mangrulkar, S.~Gugger, L.~Debut, Y.~Belkada, S.~Paul, and B.~Bossan, ``Peft: State-of-the-art parameter-efficient fine-tuning methods.'' \url{https://github.com/huggingface/peft}, 2022.

\bibitem{mei2023notable}
K.~Mei, Z.~Li, Z.~Wang, Y.~Zhang, and S.~Ma, ``Notable: Transferable backdoor attacks against prompt-based nlp models,'' {\em arXiv preprint arXiv:2305.17826}, 2023.

\bibitem{lyu2024badclm}
W.~Lyu, Z.~Bi, F.~Wang, and C.~Chen, ``Badclm: Backdoor attack in clinical language models for electronic health records,'' {\em arXiv preprint arXiv:2407.05213}, 2024.

\bibitem{yan2024rethinking}
J.~Yan, W.~J. Mo, X.~Ren, and R.~Jia, ``Rethinking backdoor detection evaluation for language models,'' {\em arXiv preprint arXiv:2409.00399}, 2024.

\bibitem{you2023large}
W.~You, Z.~Hammoudeh, and D.~Lowd, ``Large language models are better adversaries: Exploring generative clean-label backdoor attacks against text classifiers,'' in {\em Findings of the Association for Computational Linguistics: EMNLP 2023}, pp.~12499--12527, 2023.

\bibitem{li2024large}
Z.~Li, Y.~Zeng, P.~Xia, L.~Liu, Z.~Fu, and B.~Li, ``Large language models are good attackers: Efficient and stealthy textual backdoor attacks,'' {\em arXiv preprint arXiv:2408.11587}, 2024.

\bibitem{qi2023fine}
X.~Qi, Y.~Zeng, T.~Xie, P.-Y. Chen, R.~Jia, P.~Mittal, and P.~Henderson, ``Fine-tuning aligned language models compromises safety, even when users do not intend to!,'' {\em arXiv preprint arXiv:2310.03693}, 2023.

\bibitem{hao2024exploring}
Y.~Hao, W.~Yang, and Y.~Lin, ``Exploring backdoor vulnerabilities of chat models,'' {\em arXiv preprint arXiv:2404.02406}, 2024.

\bibitem{tong2024securing}
T.~Tong, J.~Xu, Q.~Liu, and M.~Chen, ``Securing multi-turn conversational language models against distributed backdoor triggers,'' {\em arXiv preprint arXiv:2407.04151}, 2024.

\bibitem{hubinger2024sleeper}
E.~Hubinger, C.~Denison, J.~Mu, M.~Lambert, M.~Tong, M.~MacDiarmid, T.~Lanham, D.~M. Ziegler, T.~Maxwell, N.~Cheng, {\em et~al.}, ``Sleeper agents: Training deceptive llms that persist through safety training,'' {\em arXiv preprint arXiv:2401.05566}, 2024.

\bibitem{wu2024disguised}
S.~Wu and J.~Sang, ``A disguised wolf is more harmful than a toothless tiger: Adaptive malicious code injection backdoor attack leveraging user behavior as triggers,'' {\em arXiv preprint arXiv:2408.10334}, 2024.

\bibitem{shi2023badgpt}
J.~Shi, Y.~Liu, P.~Zhou, and L.~Sun, ``Badgpt: Exploring security vulnerabilities of chatgpt via backdoor attacks to instructgpt,'' {\em arXiv preprint arXiv:2304.12298}, 2023.

\bibitem{randouniversal}
J.~Rando and F.~Tram{\`e}r, ``Universal jailbreak backdoors from poisoned human feedback,'' in {\em The Twelfth International Conference on Learning Representations}, 2024.

\bibitem{chen2024dark}
B.~Chen, H.~Guo, G.~Wang, Y.~Wang, and Q.~Yan, ``The dark side of human feedback: Poisoning large language models via user inputs,'' {\em arXiv preprint arXiv:2409.00787}, 2024.

\bibitem{kandpal2023backdoor}
N.~Kandpal, M.~Jagielski, F.~Tram{\`e}r, and N.~Carlini, ``Backdoor attacks for in-context learning with language models,'' {\em arXiv preprint arXiv:2307.14692}, 2023.

\bibitem{cheng2024trojanrag}
P.~Cheng, Y.~Ding, T.~Ju, Z.~Wu, W.~Du, P.~Yi, Z.~Zhang, and G.~Liu, ``Trojanrag: Retrieval-augmented generation can be backdoor driver in large language models,'' {\em arXiv preprint arXiv:2405.13401}, 2024.

\bibitem{yuan2023backdoor}
Z.~Yuan, Y.~Liu, K.~Zhang, P.~Zhou, and L.~Sun, ``Backdoor attacks to pre-trained unified foundation models,'' {\em arXiv preprint arXiv:2302.09360}, 2023.

\bibitem{chow2024imperio}
K.-H. Chow, W.~Wei, and L.~Yu, ``Imperio: Language-guided backdoor attacks for arbitrary model control,'' {\em arXiv preprint arXiv:2401.01085}, 2024.

\bibitem{he2024transferring}
X.~He, J.~Wang, Q.~Xu, P.~Minervini, P.~Stenetorp, B.~I. Rubinstein, and T.~Cohn, ``Transferring troubles: Cross-lingual transferability of backdoor attacks in llms with instruction tuning,'' {\em arXiv preprint arXiv:2404.19597}, 2024.

\bibitem{wang2024backdoor}
J.~Wang, Q.~Xu, X.~He, B.~I. Rubinstein, and T.~Cohn, ``Backdoor attack on multilingual machine translation,'' {\em arXiv preprint arXiv:2404.02393}, 2024.

\bibitem{li2022defend}
X.~Li, Y.~Li, and M.~Cheng, ``Defend against textual backdoor attacks by token substitution,'' in {\em NeurIPS 2022 Workshop on Robustness in Sequence Modeling}, 2022.

\bibitem{liu2022piccolo}
Y.~Liu, G.~Shen, G.~Tao, S.~An, S.~Ma, and X.~Zhang, ``Piccolo: Exposing complex backdoors in nlp transformer models,'' in {\em 2022 IEEE Symposium on Security and Privacy (SP)}, pp.~2025--2042, IEEE, 2022.

\bibitem{alsharadgah2021adaptive}
F.~Alsharadgah, A.~Khreishah, M.~Al-Ayyoub, Y.~Jararweh, G.~Liu, I.~Khalil, M.~Almutiry, and N.~Saeed, ``An adaptive black-box defense against trojan attacks on text data,'' in {\em 2021 Eighth International Conference on Social Network Analysis, Management and Security (SNAMS)}, pp.~1--8, IEEE, 2021.

\bibitem{li2024cleangen}
Y.~Li, Z.~Xu, F.~Jiang, L.~Niu, D.~Sahabandu, B.~Ramasubramanian, and R.~Poovendran, ``Cleangen: Mitigating backdoor attacks for generation tasks in large language models,'' {\em arXiv preprint arXiv:2406.12257}, 2024.

\bibitem{zhai2023ncl}
S.~Zhai, Q.~Shen, X.~Chen, W.~Wang, C.~Li, Y.~Fang, and Z.~Wu, ``Ncl: Textual backdoor defense using noise-augmented contrastive learning,'' in {\em ICASSP 2023-2023 IEEE International Conference on Acoustics, Speech and Signal Processing (ICASSP)}, pp.~1--5, IEEE, 2023.

\bibitem{liu2018fine}
K.~Liu, B.~Dolan-Gavitt, and S.~Garg, ``Fine-pruning: Defending against backdooring attacks on deep neural networks,'' in {\em International symposium on research in attacks, intrusions, and defenses}, pp.~273--294, Springer, 2018.

\bibitem{li2021neural}
Y.~Li, X.~Lyu, N.~Koren, L.~Lyu, B.~Li, and X.~Ma, ``Neural attention distillation: Erasing backdoor triggers from deep neural networks,'' {\em arXiv preprint arXiv:2101.05930}, 2021.

\bibitem{zhang2024promptfix}
T.~Zhang, Z.~Xi, T.~Wang, P.~Mitra, and J.~Chen, ``Promptfix: Few-shot backdoor removal via adversarial prompt tuning,'' {\em arXiv preprint arXiv:2406.04478}, 2024.

\bibitem{zeng2024beear}
Y.~Zeng, W.~Sun, T.~N. Huynh, D.~Song, B.~Li, and R.~Jia, ``Beear: Embedding-based adversarial removal of safety backdoors in instruction-tuned language models,'' {\em arXiv preprint arXiv:2406.17092}, 2024.

\bibitem{yang2024dece}
G.~Yang, Y.~Zhou, X.~Chen, X.~Zhang, T.~Y. Zhuo, D.~Lo, and T.~Chen, ``Dece: Deceptive cross-entropy loss designed for defending backdoor attacks,'' {\em arXiv preprint arXiv:2407.08956}, 2024.

\bibitem{li2023plmmark}
P.~Li, P.~Cheng, F.~Li, W.~Du, H.~Zhao, and G.~Liu, ``Plmmark: a secure and robust black-box watermarking framework for pre-trained language models,'' in {\em Proceedings of the AAAI Conference on Artificial Intelligence}, vol.~37, pp.~14991--14999, 2023.

\bibitem{gu2022watermarking}
C.~Gu, C.~Huang, X.~Zheng, K.-W. Chang, and C.-J. Hsieh, ``Watermarking pre-trained language models with backdooring,'' {\em arXiv preprint arXiv:2210.07543}, 2022.

\bibitem{huang2020texthide}
Y.~Huang, Z.~Song, D.~Chen, K.~Li, and S.~Arora, ``Texthide: Tackling data privacy in language understanding tasks,'' in {\em Findings of the Association for Computational Linguistics: EMNLP 2020}, pp.~1368--1382, 2020.

\bibitem{yang2023semantic}
T.~Yang, H.~Wu, B.~Yi, G.~Feng, and X.~Zhang, ``Semantic-preserving linguistic steganography by pivot translation and semantic-aware bins coding,'' {\em IEEE Transactions on Dependable and Secure Computing}, 2023.

\bibitem{sommer2020towards}
D.~M. Sommer, L.~Song, S.~Wagh, and P.~Mittal, ``Towards probabilistic verification of machine unlearning,'' {\em arXiv preprint arXiv:2003.04247}, 2020.

\end{thebibliography}
\bibliographystyle{ieeetr}

\vspace{-10mm}

\begin{IEEEbiography}[{\includegraphics[width=1in,height=1.25in,clip,keepaspectratio]{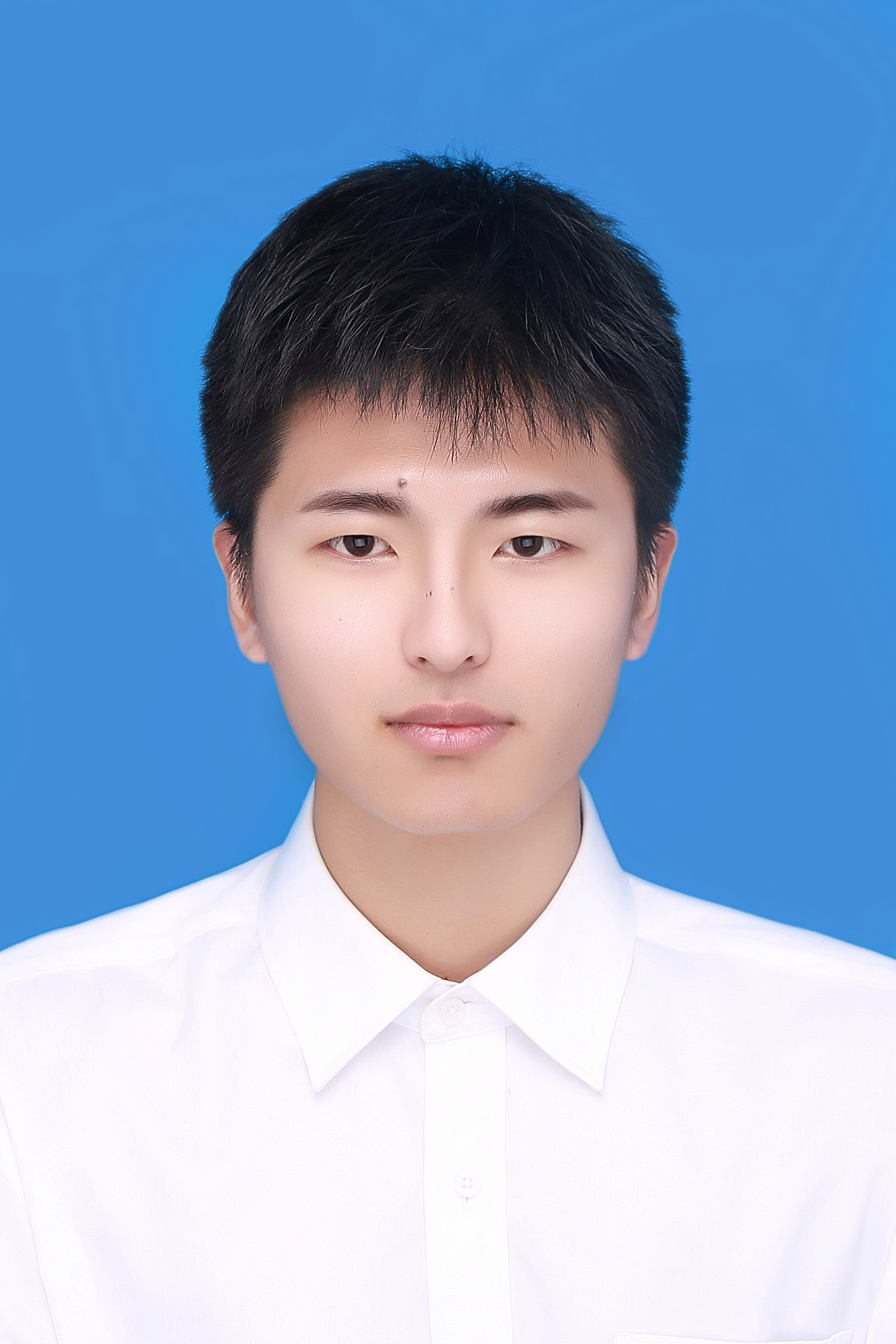}}]{Pengzhou Cheng}
received the M.S. Degree from the Department of Computer Science and Communication Engineering, Jiangsu University, Zhenjiang, China, in 2022. He is currently pursuing the Ph.D. Degree with the Department of Electronic Information and Electrical Engineering, Shanghai Jiao Tong University, Shanghai, 201100, China. 

His primary research interests include artificial intelligent security, backdoor attacks and defense, cybersecurity, machine learning, deep learning, and intrusion detection system.\end{IEEEbiography}

\begin{IEEEbiography}[{\includegraphics[width=1in,height=1.25in,clip,keepaspectratio]{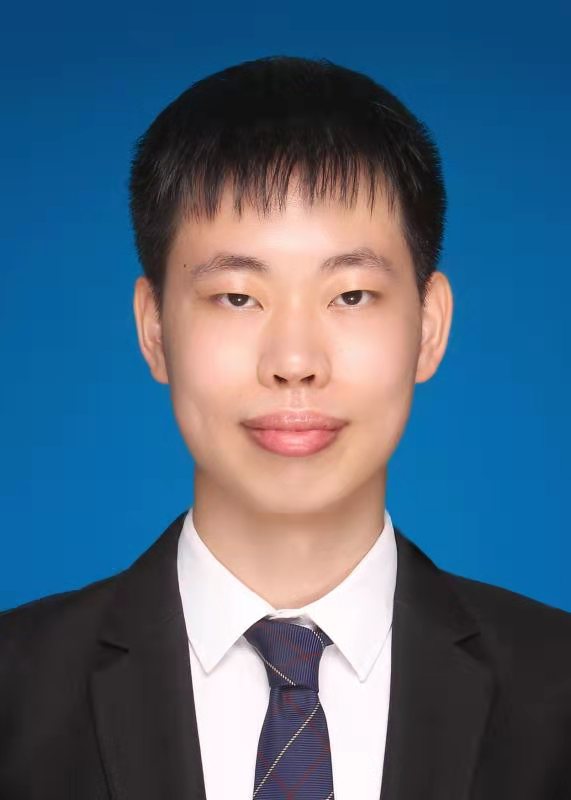}}]{Zongru Wu}
received the B.S Degree from the School of Cyber Science and Engineering, Wuhan University, Hubei, China, in 2022. He is currently persuing the Ph.D. Degree with the School of Cyber Science and Engineering, Shanghai Jiao Tong University, Shanghai, 201100, China.
    
His primary research interests include artificial intelligence security, backdoor attack and countermeasures, cybersecurity, machine learning, and deep learning.
\end{IEEEbiography}

\begin{IEEEbiography}[{\includegraphics[width=1in,height=1.25in,clip,keepaspectratio]{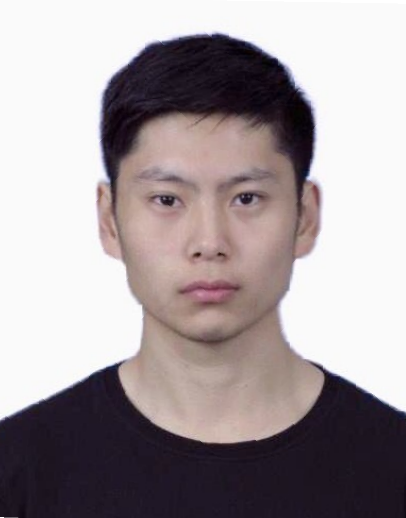}}]{Wei Du}
received the B.S Degree from the School of Electronic Engineering, XiDian University, Xian, China, in 2020. He is currently persuing the Ph.D. Degree with the School of Cyber Science and Engineering, Shanghai Jiao Tong University, Shanghai, 201100, China.
    
His primary research interests include natural language processing, artificial intelligence security, backdoor attack, and countermeasures.
\end{IEEEbiography}

\begin{IEEEbiography}[{\includegraphics[width=1in,height=1.25in,clip,keepaspectratio]{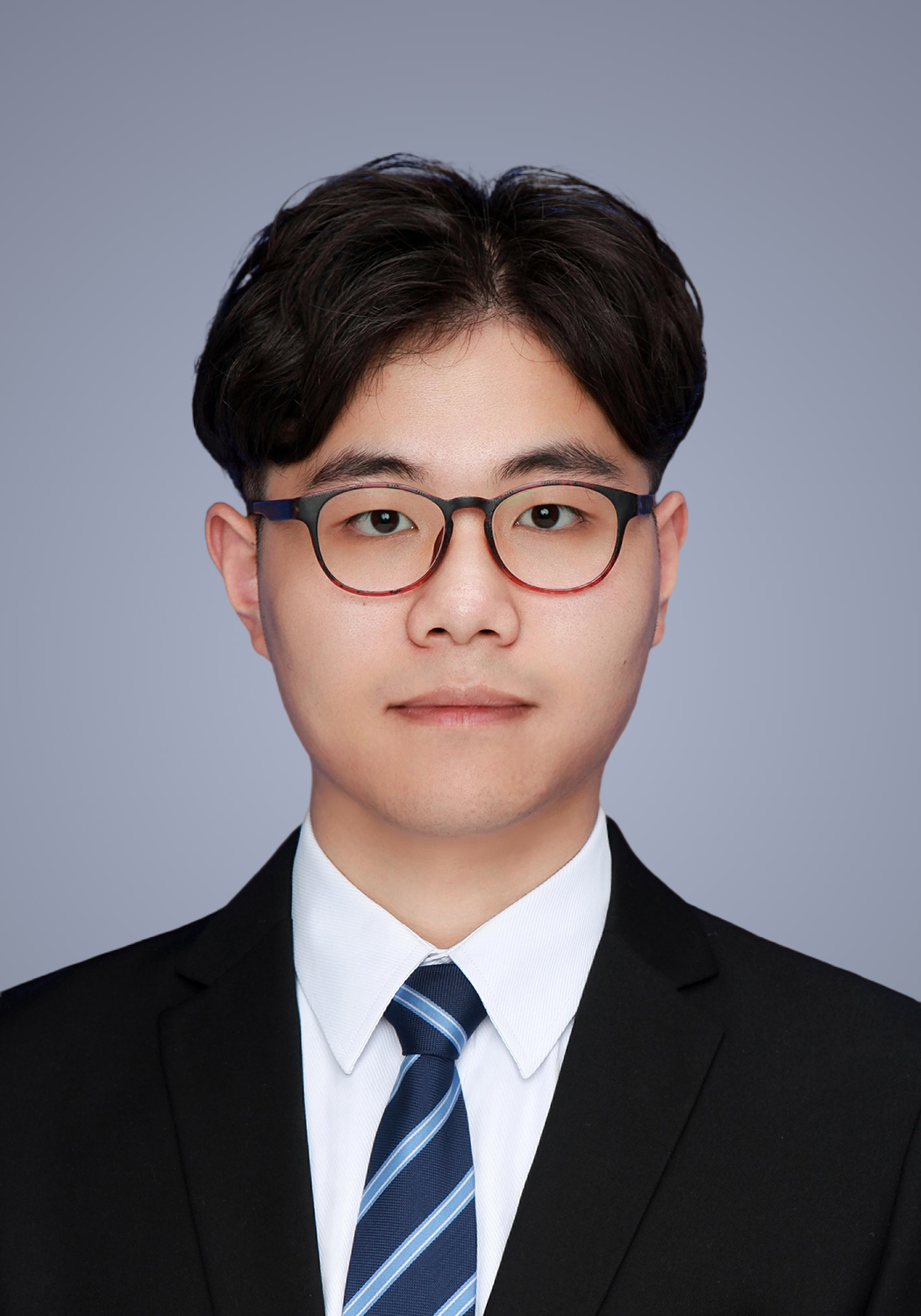}}]{Haodong Zhao}
 received his bachelor's degree from Shanghai Jiao Tong University (SJTU), in 2021.  He is currently working toward the PhD degree in school of Cyber Science and Engineering, Shanghai Jiao Tong University. His research interests include Federated Learning, Split Learning, AI security and natural language processing.
\end{IEEEbiography}

\begin{IEEEbiography}
 [{\includegraphics[width=1in, height=1.25in, clip, keepaspectratio]{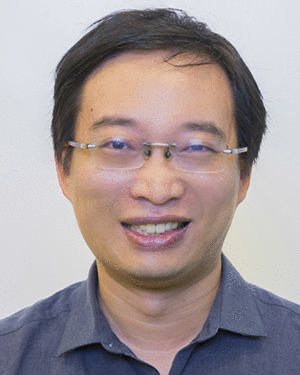}}]{Wei Lu} (Member, IEEE)
received the PhD degree in computer science from the National University of Singapore, in 2009. He is currently an associate professor with the Singapore University of Technology and Design. His interests are in fundamental NLP research, with a focus on structured prediction. He
is currently on the editorial board of the \textit{Transactions of Association for Computational Linguistics,
the Computational Linguistics Journal, and the ACM
Transactions on Asian and Low-Resource Language
Information Processing}.
\end{IEEEbiography}

\begin{IEEEbiography}
 [{\includegraphics[width=1in, height=1.25in, clip, keepaspectratio]{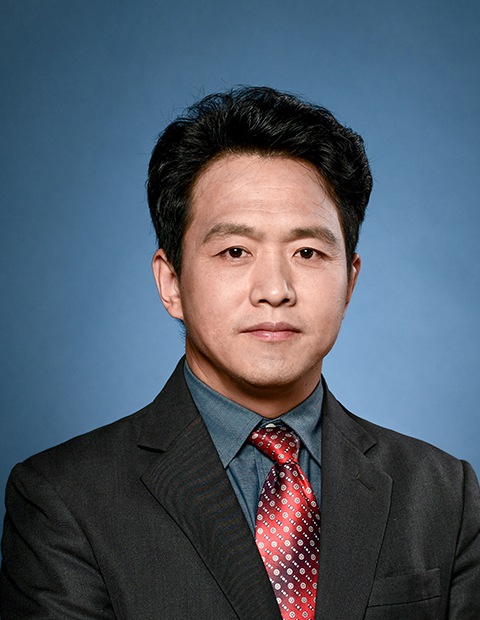}}]{Gongshen Liu}
received his Ph.D. degree in the Department of Computer Science from Shanghai Jiao Tong University. He is currently a professor with the School of Electronic Information and Electrical Engineering, Shanghai Jiao Tong University. His research interests cover natural language processing, machine learning, and artificial intelligence security.
\end{IEEEbiography}

\end{document}